\documentclass[british,english,aps,manuscript,superscriptaddress]{revtex4-1}
\usepackage[T1]{fontenc}
\usepackage[latin9]{inputenc}
\usepackage{float}
\usepackage{rotfloat}
\usepackage{amsmath}
\usepackage{amssymb}
\usepackage{graphicx}

\makeatletter

\newcommand{\lyxmathsym}[1]{\ifmmode\begingroup\def\b@ld{bold}
  \text{\ifx\math@version\b@ld\bfseries\fi#1}\endgroup\else#1\fi}

\providecommand{\tabularnewline}{\\}

\@ifundefined{textcolor}{}
{%
 \definecolor{BLACK}{gray}{0}
 \definecolor{WHITE}{gray}{1}
 \definecolor{RED}{rgb}{1,0,0}
 \definecolor{GREEN}{rgb}{0,1,0}
 \definecolor{BLUE}{rgb}{0,0,1}
 \definecolor{CYAN}{cmyk}{1,0,0,0}
 \definecolor{MAGENTA}{cmyk}{0,1,0,0}
 \definecolor{YELLOW}{cmyk}{0,0,1,0}
}

\@ifundefined{showcaptionsetup}{}{%
 \PassOptionsToPackage{caption=false}{subfig}}
\usepackage{subfig}
\makeatother

\usepackage{babel}
\begin{document}

\title{A multi-term solution of the space-time Boltzmann equation for electrons
in gaseous and liquid argon}

\author{G. J. Boyle}

\affiliation{College of Science, Technology \& Engineering, James Cook University,
Townsville, QLD 4810, Australia}

\author{W. J. Tattersall}

\affiliation{College of Science, Technology \& Engineering, James Cook University,
Townsville, QLD 4810, Australia}

\affiliation{Research School of Physical Sciences and Engineering, Australian
National University, Canberra, ACT 0200, Australia}

\author{D. G. Cocks}

\affiliation{College of Science, Technology \& Engineering, James Cook University,
Townsville, QLD 4810, Australia}

\author{R. P. McEachran}

\affiliation{Research School of Physical Sciences and Engineering, Australian
National University, Canberra, ACT 0200, Australia}

\author{R. D. White}

\affiliation{College of Science, Technology \& Engineering, James Cook University,
Townsville, QLD 4810, Australia}
\begin{abstract}
In a recent paper \cite{Boyletal15} the scattering and transport
of excess electrons in liquid argon in the hydrodynamic regime was
investigated, generalizing the seminal works of Lekner and Cohen \cite{Lekn67,CoheLekn67}
with modern scattering theory techniques and kinetic theory. In this
paper, the discussion is extended to the non-hydrodynamic regime through
the development of a full multi-term space-time solution of Boltzmann's
equation for electron transport in gases and liquids using a novel
operator-splitting method. A Green's function formalism is considered
that enables flexible adaptation to various experimental systems.
The spatio-temporal evolution of electrons in liquids in the hydrodynamic
regime is studied for a benchmark model Percus-Yevick liquid as well
as for liquid argon. The temporal evolution of Franck-Hertz oscillations
are observed for liquids, with striking differences in the spatio-temporal
development of the velocity distribution function components between
the uncorrelated gas and true liquid approximations in argon. Transport
properties calculated from the non-hydrodynamic theory in the long
time limit, and under steady-state Townsend conditions, are benchmarked
against hydrodynamic transport coefficients. 
\end{abstract}
\maketitle

\section{Introduction\label{sec:Introduction}}

The study of electron transport in dense gases and liquids is of interest
in understanding the fundamental microscopic scattering processes
involved, and to technological applications including liquid-state
electronics \cite{schmidt1997liquid}, high-energy particle detectors
\cite{Marc13,Aprile2010g,Aprile2011g,Sumner2005g}, plasma medicine
\foreignlanguage{british}{\cite{Kong2009,Tian2014,Norberg2014,Chen2014}}
and radiation dosimetry \cite{Zhenetal06,Munoetal12,Whitetal12,Nikjoo2008}.
For these technologies to reach their full potential requires a detailed
understanding of the full spatio-temporal behaviour of electrons in
dense gases, liquids, and other bio-structures, typically under non-equilibrium
conditions. 

In a recent paper \cite{Boyletal15} the transport of excess electrons
in liquid argon was considered from ab initio liquid phase cross-sections
calculated using the Dirac-Fock scattering equations. The approach
detailed in the seminal works by Lekner and Cohen \cite{Lekn67,CoheLekn67}
has been revisited with modern scattering theory techniques, where
the original treatment was extended to consider multipole polarizabilities
and a non-local treatment of exchange. With an increase in density,
several important density effects become significant, most notably
(i) the coherent scattering from multiple scattering centres, (ii)
the screening of the long range polarization potential due to induced
multipoles in the bulk, and (iii) the contribution of the bulk to
the effective potential experienced by the electron. Transport coefficients
such as drift velocities and characteristic energies calculated in
the hydrodynamic regime with our hydrodynamic multi-term Boltzmann
equation solution were in good agreement with swarm experiment measurements
in both gas- and liquid-phase argon \cite{Boyle2015a}. In this work
we extend the discussion to an investigation of liquid state in the
non-hydrodynamic regime, using the same electron-argon potentials
and cross-sections presented in \cite{Boyletal15}.

The solution of the the full temporal-, spatial- and energy-dependent
Boltzmann equation is formidable, both mathematically and computationally.
Historically, the majority of kinetic theory investigations have focused
on the hydrodynamic regime where spatial gradients are small, and
have considered increasingly complex space- and time-dependent hydrodynamic
behaviours and field configurations (see reviews \cite{RobsNess86,Whitetal99,White2002b,Whitetal09,LoffSige09}).
In situations where the hydrodynamic regime is not applicable, the
space-time dependence of the phase space distribution function cannot
be projected onto the number density and a density gradient expansion
is no longer valid. Instead the configuration-space dependence of
the Boltzmann equation must be treated on equal footing with the energy-space
dependence, which makes for a difficult problem even for simple geometries
\cite{Lietal02,Poroetal05a,Poroetal05b}. It is no surprise that systematic
studies of non-hydrodynamic phenomena lag behind their hydrodynamic
counterparts. The prototypical example of non-hydrodynamic phenomena
is the Frank-Hertz experiment \cite{FranHert14,Robsetal00}, which
helped lay the foundations for quantum and atomic physics. Extensive
theoretical studies of non-hydrodynamic electron phenomena have been
performed including field free spatial relaxation \cite{WinkSige01},
and spatial relaxation in the presence of uniform \cite{Winketal96,SigeWink97b,SigeWink97,Winketal97},
non-uniform \cite{Loffetal03} and periodic electric fields \cite{Sigeetal98,SigeWink00,Sigeetal00}.
Similar kinetic studies on the spatial relaxation of electrons in
uniform and spatially periodic fields have been performed by Golubovskii
et al. \cite{Goluetal00,Goluetal02,Goluetal98,Goluetal99}. Li and
co-workers have considered arbitrary electric and magnetic field configurations
with a multi-term analysis \cite{Whitetal99,Lietal06,Dujko2011}.
Solution of the Boltzmann equation for electrons including both the
space and time dependence have also recently been performed \cite{GoedMeij93,MahmYous97,LoffWink01},
however these authors restricted their calculations to a two-term
approximation in Legendre polynomials in order to make the problem
computationally feasible. Limitations of the two-term approximation
for molecular gases are well known \cite{White2003a}. Prior to \cite{Boyletal15},
all studies of electron transport in liquids were in the hydrodynamic
or spatially homogeneous regimes, and restricted to the two-term approximation. 

In this study, we present a full multi-term space-time dependent solution
of Boltzmann's equation, capable of handling highly non-equilibrium
electron transport in dilute gases, dense gases and liquids under
non-hydrodynamic conditions. To our knowledge, this is the first time
such a complete solution of Boltzmann's equation has been developed.
In addition, by solving for the spatio-temporal evolution of the Boltzmann
equation Green's function, the technique is quite general in its application,
enabling various experimental configurations (temporal and spatial
initial and boundary conditions) and practical devices to be modelled
from a single solution. This work extends the Boltzmann equation framework
to applications and accuracies comparable to those achieved using
the Monte-Carlo simulations of Petrovic, Dujko and co-workers \cite{Dujketal08,Petrovic2013,Dujko2014}. 

We begin the paper in Section \ref{sec:Theory} with a brief overview
of the multi-term solution of Boltzmann's equation for electrons in
structured materials, such as liquids. We then detail our operator
splitting treatment of the space, time and energy dependence in Section
\ref{sec:Solution-technique}. In Section \ref{sec:Modified-Percus-Yevick-Hard-Sphe}
we present solutions for a model hard-sphere liquid system with a
Percus-Yevick structure factor used to simulate a prototypical liquid
with realistic pair correlations. A simple inelastic channel is included
to induce periodic oscillatory structures (an idealized verion of
the well known Frank-Hertz experiment \cite{Robsetal00}) which can
act as a non-hydrodynamic benchmark. Lastly, in Section \ref{sec:Electrons-in-dilute}
we investigate the temporal and spatial evolution of the phase-space
distribution for electrons in liquid argon, using microscopic cross-sections
which have been derived previously \cite{Boyletal15}. The issues
with treating liquid systems as gaseous systems with increased density,
and the implications for various applications including liquid argon
time projection chambers, are highlighted.

\section{Theory\label{sec:Theory}}

The fundamental kinetic equation describing the evolution of an electron
swarm in a gaseous or liquid background medium subject to an electric
field, $\mathbf{E}$, is the Boltzmann equation for the phase-space
distribution function $f\equiv f\left(\mathbf{r},\mathbf{v},t\right)$
\cite{WhitRobs11}:

\begin{align}
\left(\frac{\partial}{\partial t}+\mathbf{v}\cdot\frac{\partial}{\partial\mathbf{r}}+\frac{e\mathbf{E}}{m}\cdot\frac{\partial}{\partial\mathbf{v}}\right)f & =-J\left(f\right),\label{eq:Boltz}
\end{align}
where $t$ is time, and $\mathbf{r}$, $\mathbf{v}$, $e$ and $m$
are the position, velocity, charge and mass of the electron respectively.
The collision operator $J(f)$ accounts for interactions between the
electrons and the background material, and describes the effect of
collisions on the distribution function at a fixed position and velocity.
In essence the Boltzmann equation represents an extension of the continuity
equation to phase-space. A solution of equation (\ref{eq:Boltz})
for the distribution function yields all relevant information about
the system. Macroscopic transport properties including mean energy
and drift velocity can then be found via averages over the ensemble,
as detailed in Section \ref{sub:Transport-properties}. 

The starting point for most modern solutions of the Boltzmann equation
is the decomposition of the angular part of the velocity dependence
of equation (\ref{eq:Boltz}) in terms of spherical harmonics \cite{WhitRobs11}.
If there is a single preferred direction in the system, e.g. due to
an electric field in plane parallel geometry, then the angular dependence
of the velocity component can be adequately described by a simpler
expansion in terms of Legendre polynomials. For the plane-parallel
geometry considered in this work, $f\left(\mathbf{v},\mathbf{r},t\right)\rightarrow f\left(v,z,\mu,t\right)$,
where $\mu=\hat{\mathbf{v}}\cdot\mathbf{\hat{E}}=\cos\chi$, such
that 
\begin{align}
f\left(\mathbf{v},\mathbf{r},t\right) & =\sum_{l=0}^{\infty}f_{l}\left(v,z,t\right)P_{l}(\mu),\label{eq:Legendre}
\end{align}
where $P_{l}$ is the $l$-th Legendre polynomial. Upon substituting
the expansion~(\ref{eq:Legendre}) into equation~(\ref{eq:Boltz})
and equating the coefficients of the Legendre polynomials results
in the following system of coupled partial differential equations
for the $f_{l}$: 

\begin{equation}
\frac{\partial f_{l}}{\partial t}+\sum_{p=\pm1}\Delta_{l}^{(p)}\left(\frac{2}{m}\right)^{\frac{1}{2}}\left[U^{^{1/2}}\frac{\partial}{\partial z}+eE\left(U^{^{\frac{1}{2}}}\frac{\partial}{\partial U}+p\frac{\left(l+\frac{3p+1}{2}\right)}{2}U^{^{-\frac{1}{2}}}\right)\right]f_{l+p}=-J_{l}\left(f_{l}\right),\label{eq:BoltzLeg}
\end{equation}
for $l=0,1,2,\dots,\infty$, where $U=\frac{1}{2}mv^{2}$, $J_{l}$
is the Legendre decomposition of the collision operator, and
\begin{align}
\Delta_{l}^{(+1)} & =\frac{l}{(2l-1)};\quad\Delta_{l}^{(-1)}=\frac{(l+1)}{(2l+3)}\;.
\end{align}
Equation (\ref{eq:BoltzLeg}) represents an infinite set of coupled
partial differential equations for the expansion coefficients, $f_{l}$.
In practice, one must truncate the series (\ref{eq:Legendre}) at
a sufficiently high index, $l=l_{\mathrm{max}}$. The history of charged
particle transport in gases and liquids has been dominated by the
`two-term approximation' , i.e., where only the first two terms have
been included. The assumption of quasi-isotropy necessary for the
two-term approximation is violated in many situations, particularly
when inelastic collisions are included \cite{Whitetal02} or when
higher order moments are probed \cite{Boyletal15}. Such an assumption
is not necessary in our multi-term formalism. Instead, the truncation
parameter $l_{\mathrm{max}}$ is treated as a free parameter that
is incremented until some convergence criterion is met on the distribution
function or its velocity moments. 

In order to solve equation (\ref{eq:BoltzLeg}) we require the collision
operators for all of the relevant collisional processes, and their
representations in terms of Legendre polynomials, $J_{l}$. If we
assume that the neutral background material is at rest and in thermal
equilibrium at a temperature $T_{0}$, then the collision operator
is linear in the swarm approximation. Below we detail the specific
forms of the collision operator for particle-conserving elastic and
inelastic collisions employed. A further expansion of each collision
integral with respect to the ratio of swarm particle mass to neutral
particle mass, $m/m_{0}$, has been performed. Because this ratio
is small for electrons in argon, only the leading term of this expansion
was taken into account for each collision process and in each equation
of the system (\ref{eq:BoltzLeg}). The total collision operator can
be separated for each of the different types of processes, e.g.

\begin{align}
J_{l} & =J_{l}^{\mathrm{el}}+J_{l}^{\mathrm{exc}},
\end{align}
where $J_{l}^{\mathrm{el}}$ and $J_{l}^{\mathrm{exc}}$ are the elastic
and inelastic collision operators, respectively. For dense mediums
and low electron energies, the de Broglie wavelength of the electron
is comparable to the average inter-particle spacing $\sim n_{0}^{-1/3}$.
The electron wave is then scattered coherently from multiple scattering
centres in the medium. Cohen and Lekner \cite{CoheLekn67} showed
how to account for coherent scattering using a two-term approximation,
which has since been extended to a multi-term regime \cite{WhitRobs09,WhitRobs11}.
The Legendre projections of the elastic collision operator in the
small mass ratio limit were shown to be:
\begin{equation}
J_{l}^{\mathrm{el}}\left(f_{l}\right)=\begin{cases}
-\frac{2m}{m_{0}}U^{-\frac{1}{2}}\frac{\partial}{\partial U}\left[U^{\frac{3}{2}}\nu_{1}^{\mathrm{el}}(U)\left(f_{0}+k_{b}T_{0}\frac{\partial f_{0}}{\partial U}\right)\right] & l=0,\\
\tilde{\nu}_{l}^{\mathrm{el}}(U)f_{l}(U) & l\geq1.
\end{cases}\label{eq:DavydovOp}
\end{equation}
Here, $m_{0}$ is the mass of the background neutral, $k_{b}$ is
Boltzmann's constant, and 

\begin{equation}
\nu_{l}^{\mathrm{el}}(U)=n_{0}\left(\frac{2U}{m}\right)^{\frac{1}{2}}\left(2\pi\int_{0}^{\pi}\sigma(U,\chi)\left[1-P_{l}(\cos\chi)\right]\sin\chi d\chi\right),\label{eq:collisionfreq_gas}
\end{equation}
are the usual binary collision frequencies. A collision frequency,
$\nu$, is related to the corresponding cross section, $\sigma$,
via $\nu=n_{0}\left(\frac{2U}{m}\right)^{\frac{1}{2}}\sigma\left(U\right)$.
The $\tilde{\nu}_{l}^{\mathrm{el}}(v)$ in equation (\ref{eq:DavydovOp})
are the structure-modified counterparts to $\nu_{l}^{\mathrm{el}}(U)$,
i.e.,

\begin{equation}
\tilde{\nu}_{l}^{\mathrm{el}}(U)=n_{0}\left(\frac{2U}{m}\right)^{\frac{1}{2}}\left(2\pi\int_{0}^{\pi}\Sigma(U,\chi)\left[1-P_{l}(\cos\chi)\right]\sin\chi d\chi\right),\label{eq:collisionfreqs_coh}
\end{equation}
 where $\Sigma(U,\chi)$ is the effective differential cross-section 

\begin{equation}
\Sigma(U,\chi)=\sigma(U,\chi)\; S\left(\frac{2}{\hbar}\sqrt{2mU}\sin\frac{\chi}{2}\right),
\end{equation}
which then accounts for coherent scattering effects through the static
structure factor, $S$ \cite{CoheLekn67}. At higher energies, the
de Broglie wavelength becomes much less than the inter-particle spacing
and the effects of coherent scattering are no longer important. In
this limit, the binary scattering approximation is recovered, i.e.,
$\Sigma\rightarrow\sigma$ and $\tilde{\nu}_{l}^{\mathrm{el}}\rightarrow\nu_{l}^{\mathrm{el}}$.
$\nu_{1}^{\mathrm{el}}$ is more commonly known as the momentum transfer
collision frequency, $\nu_{m}$, which is associated with the momentum
transfer cross section, $\sigma_{m}$. Similarly, $\tilde{\nu}_{m}=\tilde{\nu}_{1}^{\mathrm{el}}$
is known as the effective momentum transfer collision frequency. We
will adopt this convention in the following discussions.

At higher fields, incoherent inelastic scattering effects, such as
electronic excitations, need to be considered \cite{WhitRobs09,WhitRobs11}.
By considering only a single inelastic channel, and assuming neutral
particles are in the ground state, the excitation collision operator
is

\begin{align}
J_{l}^{\mathrm{exc}}\left(f_{l}\right) & =\nu^{\mathrm{exc}}\left(U\right)f_{l}(U)-\begin{cases}
\begin{array}{c}
\left(\frac{U+U_{I}}{U}\right)^{\frac{1}{2}}\nu^{\mathrm{exc}}\left(U+U_{I}\right)f_{l}\left(U+U_{I}\right)\mbox{,}\\
0\mbox{,}
\end{array} & \begin{array}{c}
l=0,\\
l\geq1.
\end{array}\end{cases}\label{eq:inelastic}
\end{align}
where $U_{I}$ is the threshold energy associated with the excitation
collision frequency $\nu^{\mathrm{exc}}$.

\section{Solution technique\label{sec:Solution-technique}}

Boltzmann's equation is a non-linear integro-differential equation
involving three spatial dimensions, three velocity dimensions and
time. The Boltzmann equation consists of two parts; an advective component
(in phase space) and a component representing collisions. It is a
formidable task to solve the Boltzmann equation numerically, using
a single numerical scheme for both components and a single time-stepping
method. Because of the complexity, we choose to replace the task of
solving the full Boltzmann equation by the task of solving the configuration-space
transport, the energy-space transport and the contributions due to
collisions separately, then combining the results in a manner that
appropriately approximates the full solution. This can be achieved
via the technique known as operator splitting \foreignlanguage{british}{\cite{VerwSpor98}}.

To this end, the Legendre polynomial expansion of Boltzmann's equation
in plane parallel geometry given in equation (\ref{eq:BoltzLeg})
can be represented as
\begin{eqnarray}
\frac{\partial f_{l}}{\partial t}+S_{Z}(f_{l})+S_{U}(f_{l}) & = & 0,
\end{eqnarray}
where 
\begin{eqnarray}
S_{Z}(f_{l}) & = & \sum_{p=\pm1}\Delta_{l}^{(p)}\left(\frac{2}{m}\right)^{1/2}U^{^{1/2}}\frac{\partial}{\partial z}f_{l+p},\label{eq:OpSplitConfig}\\
S_{U}(f_{l}) & = & \sum_{p=\pm1}\Delta_{l}^{(p)}\left(\frac{2}{m}\right)^{1/2}eE\left(U^{^{1/2}}\frac{\partial}{\partial U}+p\frac{\left(l+\frac{3p+1}{2}\right)}{2}U^{^{-1/2}}\right)f_{l+p}+J_{l}\left(f_{l}\right).\label{eq:OpSplitEnergy}
\end{eqnarray}

\subsection{Operator splitting\label{sub:Operator-splitting}}

The simplest method of operator splitting, and the method employed
in this paper, is Lie-Trotter splitting \cite{BagrGodu57,Stra68},
which employs two separate operators, e.g. $S_{Z}$ and $S_{U}$,
in a sequential order. If
\begin{equation}
\frac{\partial f}{\partial t}+S_{Z}(f)+S_{U}(f)=0,
\end{equation}
then the Lie-Trotter algorithm is
\begin{align}
\frac{\partial f^{*}}{\partial t}+S_{Z}(f^{*})=0, & \mbox{\,\ with }t\in\left[t^{n},t^{n+1}\right]\mbox{ and }f^{*}\left(t^{n}\right)=f\left(t^{n}\right),\label{eq:LieTrotter}\\
\frac{\partial f^{\#}}{\partial t}+S_{U}(f^{\#})=0, & \mbox{ \,\ with }t\in\left[t^{n},t^{n+1}\right]\mbox{ and }f^{\#}\left(t^{n}\right)=f^{*}\left(t^{n+1}\right),\label{eq:LieTrotter2}
\end{align}
so that $f(t^{n+1})=f^{\#}\left(t^{n+1}\right)$, where $t^{n}$ and
$t^{n+1}$ are successive times. This simple method can be shown to
be only accurate to first order in time, and there are many other
methods available that offer higher order accuracy and often include
additional advantageous properties \cite{Stra68,DiaScha96,Ohwa98,GeriWein03,PareRuss00,PareRuss05,GottShu98}.
The major reason for this particular choice of operator splitting
algorithm is that if $S_{Z}$ is treated in an explicit manner, and
$S_{U}$ is treated in an implicit manner, then the result is essentially
the Douglas class of the Alternating Direction Implicit schemes \cite{Doug62,DougGunn64},
which is particularly successful at accurately capturing the steady-state
solution. Accurately and consistently determining the steady-state
solution can be a problem for general operator splitting methods \cite{Hund02}.

The isolation of the configuration-space dependence to the operator
$S_{Z}$ makes this particular scheme an example of dimensional splitting.
We can now investigate how we treat the configuration-space advection
and energy-space advection and collision components numerically in
detail.

\subsection{Configuration-space advection\label{sub:Configuration-space-advection}}

The operator involving the configuration-space dependence, $S_{z}$,
is given by equation (\ref{eq:OpSplitConfig}), which represents a
coupled homogeneous advection equation. As there are no derivatives
of $U$ present in $S_{z}$, the configuration-space dependence can
be solved independently for different values of $U$ which is huge
simplification when a discretization in energy space is used. The
coupled advection equation can be simplified as follows:
\begin{equation}
\frac{\partial}{\partial t}f_{l}+U^{\frac{1}{2}}\Delta_{l}^{(-)}\frac{\partial}{\partial z}f_{l-1}+U^{\frac{1}{2}}\Delta_{l}^{(+)}\frac{\partial}{\partial z}f_{l+1}=0,
\end{equation}
which can be written in matrix form,
\begin{equation}
\frac{\partial}{\partial t}\mathbf{f}+\mathbf{A}\frac{\partial}{\partial z}\mathbf{f}=\mathbf{0}
\end{equation}
where $\mathbf{f}=[f_{0},f_{1},...,f_{l_{max}}]$ and 
\begin{equation}
\mathbf{A}=U^{\frac{1}{2}}\begin{bmatrix}0 & \Delta_{0}^{(+)}\\
\Delta_{1}^{(-)} & 0 & \Delta_{1}^{(+)}\\
 & \ddots & \ddots & \ddots\\
 &  & \Delta_{l_{max}-1}^{(+)} & 0 & \Delta_{l_{max}-1}^{(+)}\\
 &  &  & \Delta_{l_{max}}^{(-)} & 0
\end{bmatrix}.
\end{equation}
By letting $\mathbf{A}=\mathbf{R}\mathbf{\Lambda}\mathbf{R^{-1}}$,
where $\mathbf{\Lambda}$ is a matrix of eigenvalues of $\mathbf{A}$
on the diagonal, and $\mathbf{R}$ are the associated eigenvectors,
then
\begin{equation}
\frac{\partial}{\partial t}\mathbf{g}+\mathbf{\mathbf{\Lambda}}\frac{\partial}{\partial z}\mathbf{g}=\mathbf{0},
\end{equation}
where $\mathbf{g}=\mathbf{R}^{-1}\mathbf{f}$, which now represents
a set of uncoupled, homogeneous advection equations. It follows from
the method of characteristics \cite{Poly02}, that
\begin{equation}
\mathbf{g}\left(t,z\right)=\mathbf{g}\left(0,z-\mathbf{\mathbf{\Lambda}}t\right).
\end{equation}
Even in this extremely simple form, the solution can be troublesome.
When discretized, the set of values $z-\mathbf{\mathbf{\Lambda}}t$
are unlikely to align with existing $z$ values, and hence some form
of interpolation is required. It can be shown that linear interpolation
is equivalent to a first order upwind finite volume method scheme
\cite{Moraetal12}. First order methods have the advantage of being
well behaved and can be used to conserve mass etc.\ with no unwanted,
unphysical oscillations, but have the disadvantage of introducing
extra numerical diffusion, particularly around regions of sharp variation
\cite{Leve07}. Higher order methods perform better at controlling
unwanted diffusion but can lead to problematic, oscillatory and unphysical
solutions. Rather than straightforward interpolation, we choose to
employ a variation of a technique well known in fluid transport, the
SHASTA algorithm of Boris and Book \cite{BoriBook73}. The SHASTA
algorithm approach, termed flux-corrected transport (FCT), leads to
a class of Eulerian finite-difference algorithms which strictly enforce
the non-negative property of realistic mass and energy densities.
As a result, steep gradients and shocks can be handled particularly
well, which is a useful property when modelling transport under non-hydrodynamic
conditions. A FCT algorithm consists conceptually of two major stages,
a transport or convective stage, followed by an anti-diffusive or
corrective stage.

We employ a simplified version of the full FCT algorithm to numerically
approximate $\mathbf{g}\left(0,z-\mathbf{\mathbf{\Lambda}}t\right)$.
Let us consider the evolution of $\mathbf{g}\left(t,z\right)$ for
a single $\mathbf{\Lambda}$, i.e., $g\left(t,z;\Lambda\right)$,
over a time interval of $\Delta t$, with a uniform configuration-space
mesh with spacing $\Delta z$. By discretizing in this way, $z_{j}=j\Delta z$
for $j=1,2\dots,n_{z}-1$, and $t_{n+1}=t_{n}+\Delta t.$ The algorithm
is as follows:
\begin{description}
\item [{1.~Shift}] The elements of $g\left(t,z;\Lambda\right)$ are shifted
to the node closest to $z-\beta$, where $\beta=\frac{\Delta t}{\Delta z}\Lambda.$
This may result in an `overshoot', but we can then propagate the shifted
solution (in step 2) either forwards or backwards in time as appropriate.
The purpose of this step is to overcome time step limitations due
to the Courant-Friedrichs-Levy (CFL) condition \cite{Couretal}, which
allows us to choose arbitrary time step sizes with respect to the
configuration-space convergence (sufficiently small time steps are
still necessary for the operator splitting accuracy etc.). By shifting
to the nearest node, the CFL condition
\begin{equation}
\left|\beta\right|=\frac{\Delta t}{\Delta z}\left|\Lambda\right|\leq1
\end{equation}
for the remaining advection is always satisfied. 
\item [{2.~Advection~with~additional~diffusion}] The advection algorithm
employed is given by
\begin{equation}
g_{j}^{n+1}=g_{j}^{n}-\frac{\beta^{\prime}}{2}\left(g_{j+1}^{n}-g_{j-1}^{n}\right)+\left(\gamma+\frac{\beta^{\prime2}}{2}\right)\left(g_{j+1}^{n}-2g_{j}^{n}+g_{j-1}^{n}\right),\label{eq:modLW}
\end{equation}
where $g_{j}^{n+1}=g\left(t_{n+1},z_{j}\right)$, and
\begin{eqnarray}
\gamma & = & \left[0,\frac{\beta^{\prime}}{2}\right],
\end{eqnarray}
is the additional numerical diffusion. The dimensionless advancement
$\beta^{\prime}=\beta-\lfloor\beta\rceil$ accounts for the shift
that has been applied in step 1. Note that $\beta^{\prime}$ may be
opposite in sign to $\beta$, which corresponds to an overshoot in
step 1. However, this does not adversely affect the procedure. If
$\gamma=0$, then equation (\ref{eq:modLW}) is the well known Lax-Wendroff
scheme \cite{Leve07}, which is accurate to second order. Historically,
the inclusion of an extra diffusion term, $\gamma$, has been used
to ensure that a density function (i.e. a function that is non-negative
by definition) remains positive, which is unconditionally enforced
everywhere if $\gamma=\frac{\beta^{\prime}}{2}$. In our case, the
$g_{j}^{n}$ include contributions from $f_{l\geq1}$, which are expected
to be negative in some regions of space. However, the presence of
$\gamma$ ensures the stability of $g_{j}^{n+1}$, which can be defined
by the requirement that $\Delta g_{j}^{n+1}<\max(\Delta g_{j-1}^{n},\Delta g_{j}^{n},\Delta g_{j+1}^{n})$
where $\Delta g_{j}^{n}=g_{j+1}^{n}-g_{j}^{n}$. When the solution
$g_{j}^{n}$ is sharply varying or, in the extreme case, a discontinuity,
the additional diffusion is necessary to suppress unphysical oscillatory
behaviour in $g_{j}^{n+1}$. 
\item [{3.~Anti-diffusion}] An `anti-diffusion' step is employed to reduce
the extra numerical diffusion introduced in (\ref{eq:modLW}) i.e.,
\begin{equation}
\bar{g}_{j}^{n+1}=g_{j}^{n+1}-\left(\gamma+\frac{\beta^{2}}{2}\right)\left(g_{j+1}^{n+1}-2g_{j}^{n+1}+g_{j-1}^{n+1}\right).\label{eq:antid}
\end{equation}
The inclusion of this extra diffusion in step 2 assures that the solution
is positive and physically realistic, and the straightforward application
of step 3 undoes this which can re-introduce a negative solution.
Boris and Book \cite{BoriBook73} suggested modifying the removal
of the erroneous diffusion by just enough to maintain positivity,
in a non-linear way (note that they worked with non-negative densities,
as we have remarked on above in step 2). This is an early example
and precursor of the modern technique of flux limiting \cite{Hart83,OsheChak84,Roe81,Zale79,Sweb84}.
In this work the full anti-diffusion step is applied in general, except
in regions where a sharp variation or discontinuity is known \textit{a
priori} (e.g. configuration-space boundaries), in which case no anti-diffusion
is applied. Unphysical oscillations can now occur, but we have found
that for the situations considered they are negligibly small. The
natural extension is to include flux limiting to prevent this unphysical
behavior but this introduces extra computational complexity. The anti-diffusion
step could also be solved implicitly rather than explicitly, but we
found that this had no significant impact on the results. 
\end{description}
It should be noted that the shift step can be performed after the
advection and anti-diffusion stages with no change in the result.
We have assumed that the boundaries are absorbing, in that the elements
of $g(t,z)$ that move outside the computational domain are lost,
and no information is introduced from outside the domain. Although
perfectly absorbing boundaries are notoriously difficult to implement
numerically, in our calculations we avoid this problem by keeping
the swarm density negligible at the simulation edges, through the
use of an adaptive mesh, see Section \ref{sub:Numerical-considerations-and}.
In practice we pre-calculate a transformation matrix (for a given
set of parameters) which combines the above three steps for each of
the grid energies.

\subsection{Energy-space advection and collisions\label{sub:Field}}

A major advantage of splitting the Boltzmann equation operator according
to equations~(\ref{eq:OpSplitConfig})-(\ref{eq:OpSplitEnergy})
is that $S_{U}$ is then the familiar, spatially homogeneous Boltzmann
equation. There is much literature on solving this equation, and we
use the approach developed previously \cite{Boyletal14a,Boyletal14b,Boyletal15,Boyletal15b}
to perform the numerical discretization and time-step. A full description
of the numerical solution of the process is given in \cite{Boyletal15b},
and we will briefly summarize it here. The equation we need to solve
is equation (\ref{eq:OpSplitEnergy}). The time dimension is discretized
with a first order implicit Euler method, which has been chosen for
its good stability properties. Similar to the work of Winkler and
collaborators \cite{Winketal84,LoffWink96,Leyhetal98}, we employ
a finite difference method to discretize the system of ODE's at centred
points using a centred difference scheme, i.e.,
\begin{align}
\left.\frac{df(U,t)}{dU}\right|_{U_{i+1/2}} & =\frac{f(U_{i+1},t)-f(U_{i},t)}{U_{i+1}-U_{i}},\\
f(U_{i+1/2},t) & =\frac{f(U_{i+1},t)+f(U_{i},t)}{2},
\end{align}
so that equation (\ref{eq:OpSplitEnergy}) evaluated at $i+1/2$ becomes,
\begin{multline}
\left.S_{U}\left(f_{l}\right)\right|_{i+1/2}=\left.J_{l}\left(f_{l}\right)\right|_{i+1/2}+\left(\frac{2}{m}\right)^{\frac{1}{2}}\sum_{p=\pm1}\Delta_{l}^{(p)}eE\left[U_{i+1/2}^{\frac{1}{2}}\left(\frac{f_{l+p}(U_{i+1},t)-f_{l+p}(U_{i},t)}{U_{i+1}-U_{i}}\right)\right.\\
\left.{}+p\frac{\left(l+\frac{3p+1}{2}\right)}{2}U_{i+1/2}^{-\frac{1}{2}}\left(\frac{f_{l+p}(U_{i+1},t)+f_{l+p}(U_{i},t)}{2}\right)\right].
\end{multline}

Although a general form can be constructed for an arbitrary grid,
the simplest case is for evenly spaced points, i.e. 
\begin{equation}
U_{i}=i\Delta U\quad\mathrm{for}\quad0\leq i\leq n_{U},
\end{equation}
where $\Delta U$ is a constant. By discretizing at the midpoint of
the two solution nodes results in a system of linear equations that
is under-determined. The extra information is naturally provided by
boundary conditions which are appended to the system.

These boundary conditions have been analyzed by Winkler and collaborators
\cite{Winketal84,LoffWink96,Leyhetal98} who investigated the multi-term,
even order approximation, and discovered that the general solution
of the steady-state hierarchy contains $\tfrac{1}{2}\left(l_{\mathrm{max}}+1\right)$
non-singular and $\tfrac{1}{2}\left(l_{\mathrm{max}}+1\right)$ singular
fundamental solutions when $U$ approaches infinity, and the physically
relevant solution has to be sought within the non-singular set of
fundamental solutions. The boundary conditions necessary for the determination
of the non-singular physically relevant solution are \cite{Winketal84}

\begin{align}
f_{l}(U=0) & =\begin{array}{cc}
0 & \mbox{for odd \ensuremath{l}},\end{array}\nonumber \\
f_{l}(U=U_{\infty}) & =\begin{array}{cc}
0 & \mbox{for even \ensuremath{l}},\end{array}\\
f_{l}(U>U_{\infty}) & =\begin{array}{cc}
0 & \mbox{for all \ensuremath{l}},\end{array}\nonumber 
\end{align}
where $U_{\infty}$ represents a sufficiently large energy. In practice,
$U_{\infty}$ has to be determined in prior calculation, and is chosen
such that the value of $f_{0}(U_{\infty})$ is less than $10^{-10}$
of the maximum value of $f_{0}$.

\subsection{Green's function solution\label{sub:Green's-function-solution}}

In our formalism and associated code, we solve for the Boltzmann equation
Green's function 

\begin{equation}
\mathcal{L}f_{l}=\delta(z-z_{0})\delta(t-t_{0}),
\end{equation}
where
\begin{equation}
\mathcal{L}f_{l}=\frac{\partial f_{l}}{\partial t}+\left(\frac{2}{m}\right)^{\frac{1}{2}}\sum_{p=\pm1}\Delta_{l}^{(p)}\left[U^{^{\frac{1}{2}}}\frac{\partial}{\partial z}+\frac{eE}{m}\left(U^{^{\frac{1}{2}}}\frac{\partial}{\partial U}+p\frac{\left(l+\frac{3p+1}{2}\right)}{2}U^{^{-\frac{1}{2}}}\right)\right]f_{l+p}+J_{l}\left(f_{l}\right)
\end{equation}
for $l=0,1,2,\dots,\infty.$ The Green's function solution, $f_{l}$,
can then be used to find the solution of the more general space-time
Boltzmann equation, i.e.
\begin{equation}
\mathcal{L}\tilde{f}_{l}=S(z,t),
\end{equation}
where $S(z,t)$ is a source term, then
\begin{equation}
\tilde{f}_{l}\left(U,z,t\right)=\int dt_{0}\ \int dz_{0}\ f_{l}\left(U,z-z_{0},t-t_{0}\right)S\left(z_{0},t_{0}\right).
\end{equation}
We do this by choosing an initial distribution in configuration-space
that is a good approximation to a delta-function, which, for this
study, is a narrow Gaussian,
\begin{equation}
\delta_{a}(z)=\frac{1}{a\sqrt{\pi}}\exp\left(-\frac{z^{2}}{a^{2}}\right),\label{eq:gauss}
\end{equation}
where $a$ is a parameter controlling the width of the Gaussian, representing
the temporal-spatial relaxation profile of a single pulse centred
on $z_{0}$ and released at $t_{0}$. In the limit of $a\rightarrow\infty$,
$\delta_{a}(z)\rightarrow\delta(z)$. The formalism is quite general,
enabling the treatment of various experiments (e.g. Pulsed Townsend
(PT), Steady-State Townsend (SST) and other drift tube configurations
\cite{HuxlCrom74} - detailed in Section \ref{sub:Transport-properties}),
as well as various source and spatial/energy space/temporal distributions,
through a single solution. This approach extends the functionality
and accuracy of Boltzmann equation solutions to those routinely achieved
by Monte Carlo simulations \cite{Dujketal05,Dujketal08,Tattetal15}.

\subsection{Numerical considerations and adaptive meshing\label{sub:Numerical-considerations-and}}

The matrix system of linear equations that result from the discretization
of the Legendre-decomposed Boltzmann equation in energy- and configuration-space
at each time step are of the size $\left(n_{z}n_{U}\left(l_{max}+1\right)\right)\times\left(n_{z}n_{U}\left(l_{max}+1\right)\right)$,
where $n_{z}$ and $n_{U}$ are the number of nodes in configuration
and energy space respectively. Due to the discretization schemes,
the matrix is sparse and sparse techniques are employed to exploit
this property. Each of these parameters are free to be increased until
some convergence criterion is met. It should be noted that, although
the two-term approximation ($l_{\mathrm{max}}=1$) has been used extensively,
it is well known that it can be insufficient in many situations \cite{White2003a}.

In order to model the spatio-temporal relaxation of a narrow Gaussian
source distribution in configuration-space with a distribution of
energies as computationally efficiently as possible, we have developed
a configuration-space node-mesh that adaptively follows the size of
the distribution throughout the simulation. In this way a small configuration-space
window is used around the original narrow Gaussian source which can
then be sufficiently resolved with a small $n_{z}$. As the initial
pulse drifts and diffuses, a small amount of information reaches and
then leaks out of the window boundaries. Before the amount of information
lost to the system exceeds some small tolerance, the window is extended
and the solution at the previous time-step calculated on the new configuration-space
mesh. We have found that the most convenient way to quantify the amount
of information on the boundary is by the relative number density,
and impose the condition that when
\begin{equation}
\int_{t_{_{0}}}^{t}dt'\ \frac{n\left(z_{L}\mbox{ or }z_{R},t'\right)}{\int dz\ n(z,t')}\geq10^{-5},
\end{equation}
then the configuration-space window is doubled (while the number of
nodes is kept the same). Here $t_{0}$ is the time of the last window
adjustment, $z_{L}$ and $z_{R}$ are locations of the left and right
configuration-space boundaries respectively. The choice to extend
the window by doubling is to make it so that the new mesh lines up
exactly with nodes of the old mesh, hence requiring no interpolation.
The accuracy of the modified Lax-Wendroff scheme used to model the
configuration-space advection \cite{Leve07} is related to the parameter
$\beta=\frac{\Delta t}{\Delta z}\Lambda$, hence by doubling $\Delta z$
after a re-adjustment, the value of $\Delta t$ can also be doubled.
This effectively allows us to use smaller time steps when our solution
is sharp and diffusing quickly, and larger time steps once the solution
has spread out and is varying less quickly. A maximum value for the
time step size still needs to be enforced however, since with bigger
time step sizes less mixing between the configuration-space and energy-space
components of the operator splitting occurs, leading to errors. 

There is one extra complication to be discussed. Since the boundaries
are absorbing, when they are re-adjusted, the number density profiles
(and distribution functions) drop directly from the built-up value
at the previous boundaries location to zero in a single $\Delta z$,
which can lead to problematic, unphysical, oscillatory solutions when
treated with the method described in Section \ref{sub:Configuration-space-advection}.
In order to combat this, we simply apply the procedure without the
final anti-diffusion step for a small amount of time on the edge and
in the newly opened regions. The extra diffusion added ensures that
the solution remains positive and give physical results, which, after
a small amount of time, ensures that the profiles decrease smoothly
to zero at the boundary. After this short correction time, we again
apply the full procedure. By not removing the added extra diffusion
we have increased the overall diffusion, but since it is only applied
for a small time and to a region where there is necessarily only a
small proportion of particles, this does not significantly affect
the transport profiles.

\subsection{Transport properties\label{sub:Transport-properties}}

The cross-sections and collision operator terms represent the microscopic
picture of electron interactions with the medium. The macroscopic
picture, e.g. transport properties that represent experimental measurables,
are obtained as averages of certain quantities with respect to the
distribution function, $f$. Among the transport properties of interest
in the current manuscript are the number density, $n$, particle flux,
$\Gamma$, and average energy, $\epsilon$, of the electron swarm,
which can be calculated via

\begin{align}
n\left(z,t\right) & =2\pi\left(\frac{2}{m}\right)^{\frac{3}{2}}\int dU\ U^{\frac{1}{2}}f_{0}(U,z,t),\label{eq:n_PT}\\
\Gamma\left(z,t\right) & =\frac{2\pi}{3}\left(\frac{2}{m}\right)^{2}\int dU\ Uf_{1}(U,z,t),\label{eq:Flux_PT}\\
\epsilon\left(z,t\right) & =\frac{1}{n\left(z,t\right)}2\pi\left(\frac{2}{m}\right)^{\frac{3}{2}}\int dU\ U^{\frac{3}{2}}f_{0}(U,z,t).\label{eq:Energy_PT}
\end{align}
Likewise, we can sample the traditional hydrodynamic transport coefficients
in this non-hydrodynamic framework, e.g. the drift velocity, $W$,
and the (longitudinal) diffusion coefficient, $D_{L}$:

\begin{equation}
W(t)=\frac{d}{dt}\left[\frac{1}{N\left(t\right)}\left(\int dz\ zn\left(z,t\right)\right)\right],\label{eq:drift}
\end{equation}

\begin{equation}
D_{L}(t)=\frac{1}{2}\frac{d}{dt}\left[\frac{1}{N\left(t\right)}\left(\int dz\ z^{2}n\left(z,t\right)\right)-\left(\frac{1}{N\left(t\right)}\int dz\ zn\left(z,t\right)\right)^{2}\right],\label{eq:diffusion}
\end{equation}
where $N(t)$ is the total number of particles: 
\begin{equation}
N\left(t\right)=\int dz\ n\left(z,t\right).
\end{equation}
When the above properties are calculated from the Green's function
solution, which corresponds to a simulation of a PT experiment, then
the transport properties for other experimental systems can also be
calculated in a straightforward manner. In this work we are also interested
in the results of a SST simulation, for which there have been previous
calculations performed for benchmark systems. Similar to \cite{Robsetal00,Lietal02,Dujketal08},
the SST transport properties can be determined from the Green's function
transport properties via

\begin{eqnarray}
f_{l}^{SST}\left(U,z\right) & = & \int_{0}^{\infty}dt_{0}\ f_{l}\left(U,z,t_{0}\right),\\
n_{SST}(z) & = & \int_{0}^{\infty}dt_{0}\ n\left(z,t_{0}\right),\label{eq:nSST}\\
\Gamma_{SST}\left(z\right) & = & \int_{0}^{\infty}dt_{0}\ n\left(z,t_{0}\right)v_{z}\left(z,t_{0}\right),\label{eq:fluxSST}\\
\epsilon_{SST}\left(z\right) & = & \frac{1}{n_{SST}(z)}\int_{0}^{\infty}dt_{0}\ n\left(z,t_{0}\right)\epsilon\left(z,t_{0}\right).\label{eq:energySST}
\end{eqnarray}
In practice the upper limit of the integrals is not $\infty$, but
a sufficiently long time for the SST transport properties to have
converged over the $z$ range considered.

\subsection{Reduced variables}

Henceforth, it is convenient to work with rescaled reduced variables.
In particular, the space and time variations will be presented as
functions of 
\begin{eqnarray}
z^{*} & = & n_{0}\sigma_{0}z,\\
t^{*} & = & n_{0}\sigma_{0}\sqrt{\frac{2e}{m}}t,
\end{eqnarray}
where \foreignlanguage{british}{$\sigma_{0}=10^{-20}\ \mathrm{m}^{2}.$}
Likewise, the electric field dependence arises through the reduced
electric field $E/n_{0}$ in units of Townsend (1 Td = 10$^{-21}$Vm$^{2}$).
By presenting results in this manner scales out the $n_{0}$ dependence,
and hence allows comparisons between the dilute gas phase and the
liquid/dense gas phase, to give a true reflection of the impact of
coherent and other scattering effects.

\section{Electron transport in a modified Percus-Yevick hard-sphere benchmark
liquid model\label{sec:Modified-Percus-Yevick-Hard-Sphe}}

In order to investigate the effects of medium structure on charged
particle transport, a model for the structure function is required.
One such model, frequently employed in the literature, is that of
a structure for a system of hard-sphere potentials obtained by applying
the Percus-Yevick approximation as a closure to the Ornstein-Zernike
equation, which yields a pair-correlation function \cite{Wert63,Thei63}.
The static structure factor is found via the Fourier transform of
the pair-correlation function, the angle-integration of which is used
directly in the numerical simulations. In particular, we use the model
of Percus and Yevick with the Verlet-Weiss correction \cite{VerlWeis71,MegePuse91}
to better emulate the structure of a real liquid (see \cite{Tattetal15}
for details). The volume fraction parameter, $\Phi$, specifies how
tightly packed the hard spheres in the medium are. It can be written
in terms of the hard-sphere radius $r$ and the neutral number density,
$n_{0}$, as $\Phi=\frac{4}{3}\pi r^{3}n_{0}$. Smaller volume fractions
indicate a larger interparticle spacing, and vice versa. We have modeled
systems with a range of densities, from $\Phi\approx0$, which approximates
a dilute gas, to $\Phi=0.4$, which states that $40\%$ of the volume
is excluded by hard-sphere potentials of the neutral molecules. Differences
in the results highlight the importance of coherent elastic scattering.

\begin{figure}[H]
\begin{centering}
\includegraphics[width=0.5\textwidth]{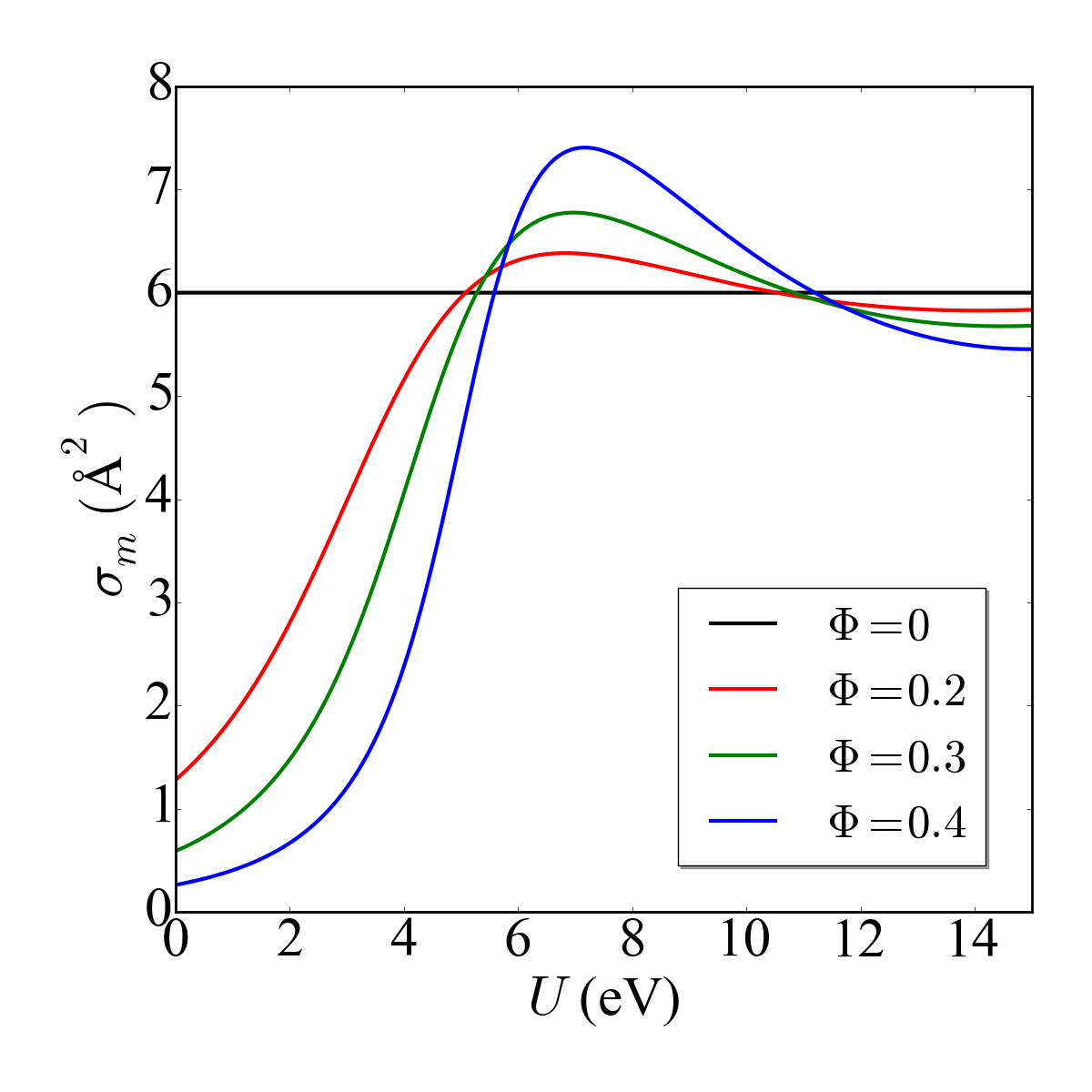}
\par\end{centering}

\selectlanguage{english}%
\centering{}\protect\caption{The variation of the elastic momentum-transfer cross-section including
structure with energy for model (\ref{eq:HSmodel}) for various volume
fractions, $\Phi$.\foreignlanguage{british}{\label{fig:PY-1}}}
\selectlanguage{british}%
\end{figure}

The remaining details required of the benchmark hard-sphere model
implemented for electron sized particles are
\begin{eqnarray}
\sigma_{m} & = & 6\ \mbox{\ensuremath{\lyxmathsym{\AA}^{2}}},\nonumber \\
\sigma^{\mathrm{exc}} & = & \begin{cases}
\begin{array}[t]{c}
0,\\
0.1\ \mbox{\ensuremath{\lyxmathsym{\AA}^{2}}},
\end{array} & \begin{array}[t]{c}
U<2\ \mbox{eV}\\
U\geq2\ \mbox{eV}
\end{array}\end{cases}\nonumber \\
\Phi & = & 0,\ 0.2,\ 0.3,\ 0.4,\nonumber \\
E/n_{0} & = & 3\ \mbox{Td},\label{eq:HSmodel}\\
m_{0} & = & 4\ \mbox{amu},\nonumber \\
T_{0} & = & 0\ \mbox{K}.\nonumber 
\end{eqnarray}
A step-like inelastic process has been included in addition to the
standard Percus-Yevick hard sphere benchmark system in model (\ref{eq:HSmodel}).
The inelastic channel introduces a periodic oscillatory non-hydrodynamic
behaviour, similar to those observed in the well-known Frank-Hertz
experiment, and can hence determine whether the numerical code is
accurately capturing the non-hydrodynamic phenomena. Figure \ref{fig:PY-1}
highlights the variation of the momentum transfer cross-section with
$\Phi$, evaluated using the cross-section in (\ref{eq:HSmodel})
together with the static structure factor from \cite{Tattetal15}.
At high energies, coherent scattering effects are suppressed, and
the various momentum transfer cross sections converge on the dilute
gas case (corresponding to $\Phi=0$) .

\selectlanguage{british}%

\selectlanguage{english}%
The source distribution is given by
\begin{equation}
f\left(U,z,0\right)=Af_{U}(U)f_{z}(z),
\end{equation}
where $f_{z}(z)$ is a narrow Gaussian in configuration-space, i.e.,
\begin{equation}
f_{z}(z)=\frac{1}{\Delta z_{0}\sqrt{2\pi}}\exp\left(-\frac{1}{2}\left(\frac{z}{\Delta z_{0}}\right)^{2}\right),
\end{equation}
(we take $\Delta z_{0}=0.1$), while $f_{U}(U)$ corresponds to drifted
Maxwellian distribution with $T=10^{5}$~K, and $\mathbf{W}=10^{4}$~ms$^{-1}$$\hat{\mathbf{E}}$,
i.e.,\foreignlanguage{british}{
\begin{equation}
f(\mathbf{v})=n\left(\frac{m}{2\pi kT}\right)^{(3/2)}\exp\left[-\frac{m}{2k_{\mathrm{b}}T}\left(\mathbf{v}-\mathbf{W}\right)^{2}\right]\thinspace,
\end{equation}
}and $A$ is a normalization constant such that $\int U^{1/2}f\left(U,z,0\right)dU=1.$

\subsection{Transport coefficients in the long-time limit}

The asymptotic values of the drift velocity and diffusion coefficient
calculated from the spatial moments (\ref{eq:drift}) and (\ref{eq:diffusion})
respectively using the full non-hydrodynamic code are displayed in
Table \ref{tab:TransCompare} for various volume fractions. Here we
compare these values with those determined from a purely hydrodynamic
formalism and associated code \cite{Boyletal15,Boyletal15b}. The
first order hydrodynamic transport coefficients, i.e., the mean energy
and drift velocity, agree to within $0.2\%$ with the asymptotic non-hydrodynamic
values for the volume fractions considered. The hydrodynamic and non-hydrodynamic
calculations of the longitudinal diffusion coefficient agree to within
$0.7\%$. As the volume fraction increases, both the mean energy,
drift velocity and diffusion coefficient increase monotonically, a
consequence of the coherent scattering, where at low energies, increasing
volume fractions leads to decreasing structure factors at low $\Delta k$,
and hence decreased momentum-transfer cross-sections. For further
discussion on the physical variation of the hydrodynamic transport
coefficients with volume fraction the reader is referred to \cite{WhitRobs11,Boyletal12,Tattetal15}.

\begin{table}[H]
\protect\caption{Comparison of the transport quantities calculated from non-hydrodynamic
(first row) and time asymptotic hydrodynamic (second row) formalisms
for model (\ref{eq:HSmodel}) at various volume fractions $\Phi$.\label{tab:TransCompare}}

\centering{}%
\begin{tabular}{cccc}
\quad{}$\Phi$\quad{} & \quad{}\quad{}$\epsilon$ \quad{}\quad{} & \quad{}\quad{}$W$ \quad{}\quad{} & \quad{}\quad{}$n_{0}D_{L}$\quad{}\quad{}\tabularnewline
 & {[}eV{]} & {[}$10^{4}\,\textrm{m\ensuremath{s^{-1}}}${]} & {[}$10^{24}\,\textrm{m\ensuremath{^{-1}}\ensuremath{s^{-1}}}${]}\tabularnewline
\hline 
\hline 
0 & 0.8335 & 1.385 & 2.386\tabularnewline
 & 0.8337 & 1.385 & 2.387\tabularnewline
\hline 
0.2 & 0.9765 & 3.397 & 6.333\tabularnewline
 & 0.9772 & 3.391 & 6.328\tabularnewline
\hline 
0.3 & 1.080 & 5.929 & 11.22\tabularnewline
 & 1.080 & 5.921 & 11.24\tabularnewline
\hline 
0.4 & 1.233 & 10.52 & 19.51\tabularnewline
 & 1.233 & 10.51 & 19.63\tabularnewline
\hline 
\end{tabular}
\end{table}

\subsection{Space-time evolution of the phase-space distribution and its velocity
moments\label{sub:Space-time-evolutionPY}}

\begin{sidewaysfigure}
\begin{minipage}[t]{1\columnwidth}%
\subfloat{\includegraphics[width=0.25\textwidth]{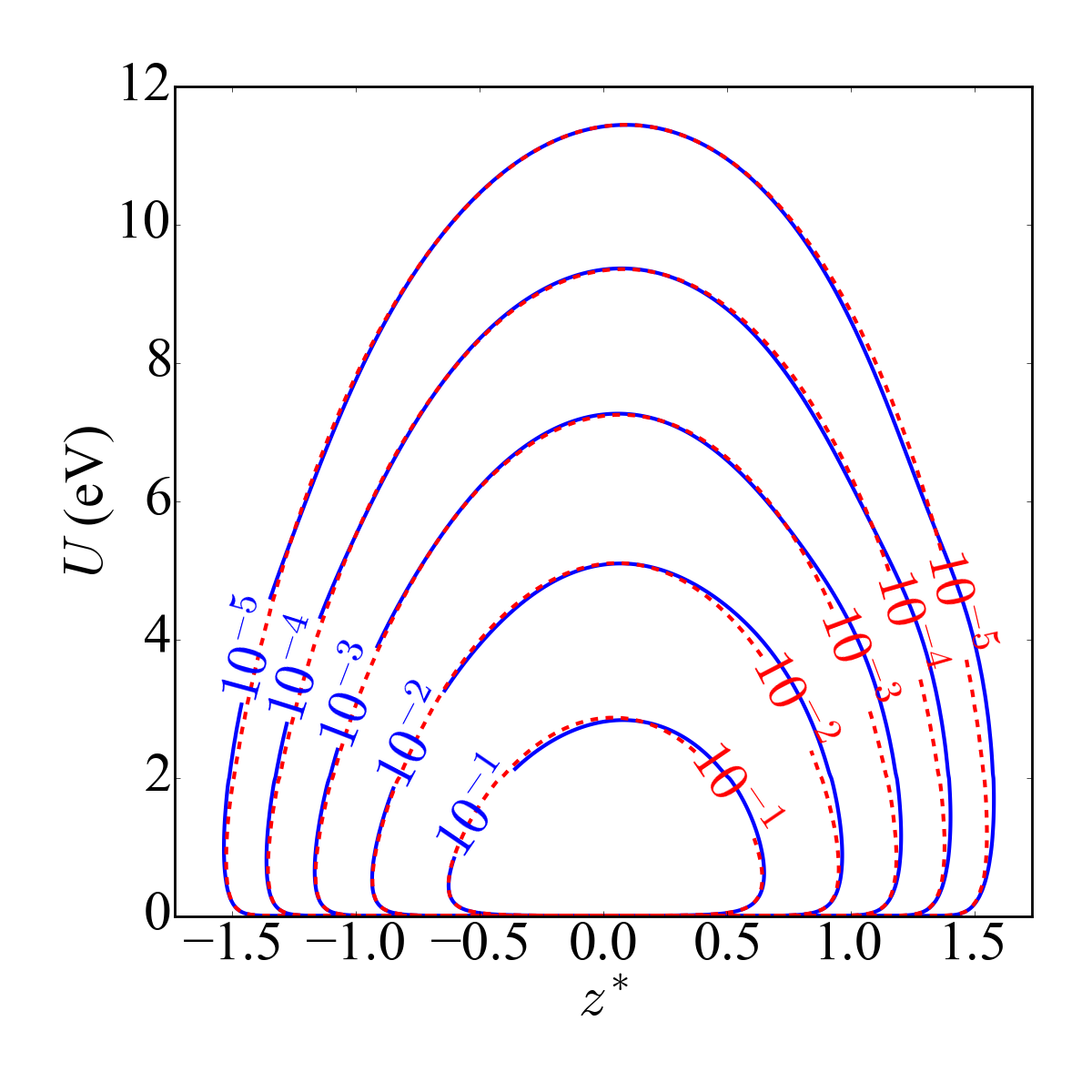}}\subfloat{\includegraphics[width=0.25\textwidth]{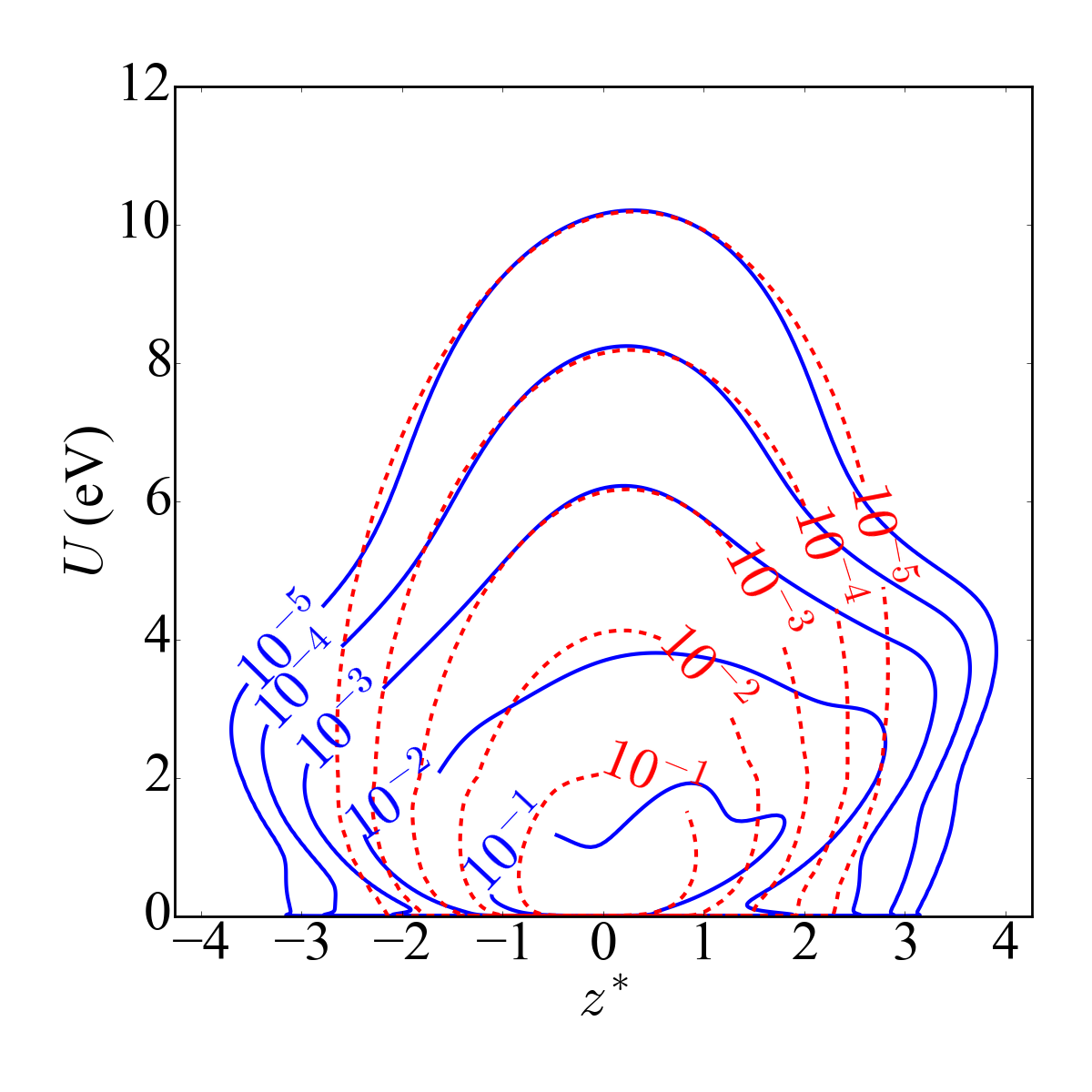}}\subfloat{\includegraphics[width=0.25\textwidth]{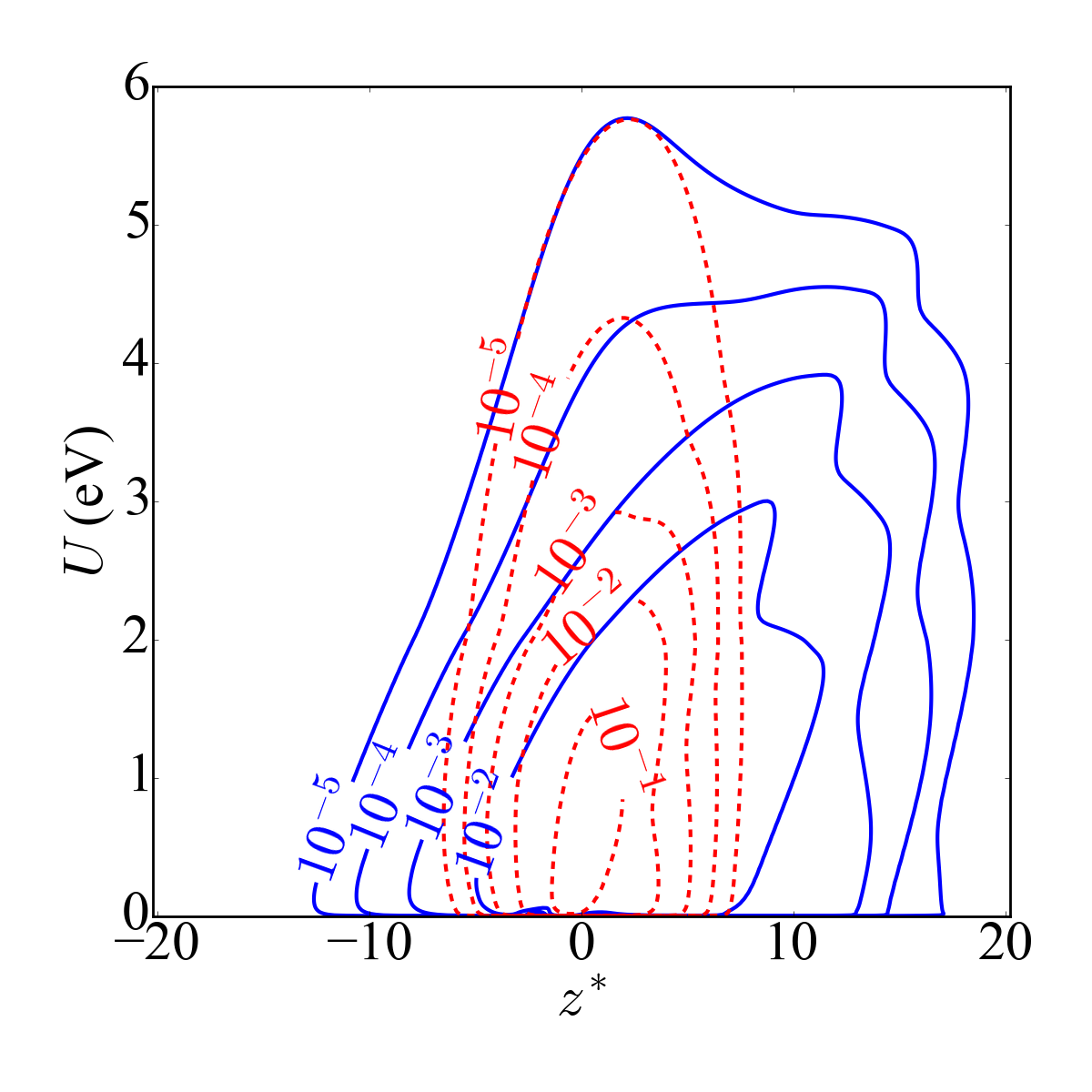}}\subfloat{\includegraphics[width=0.25\textwidth]{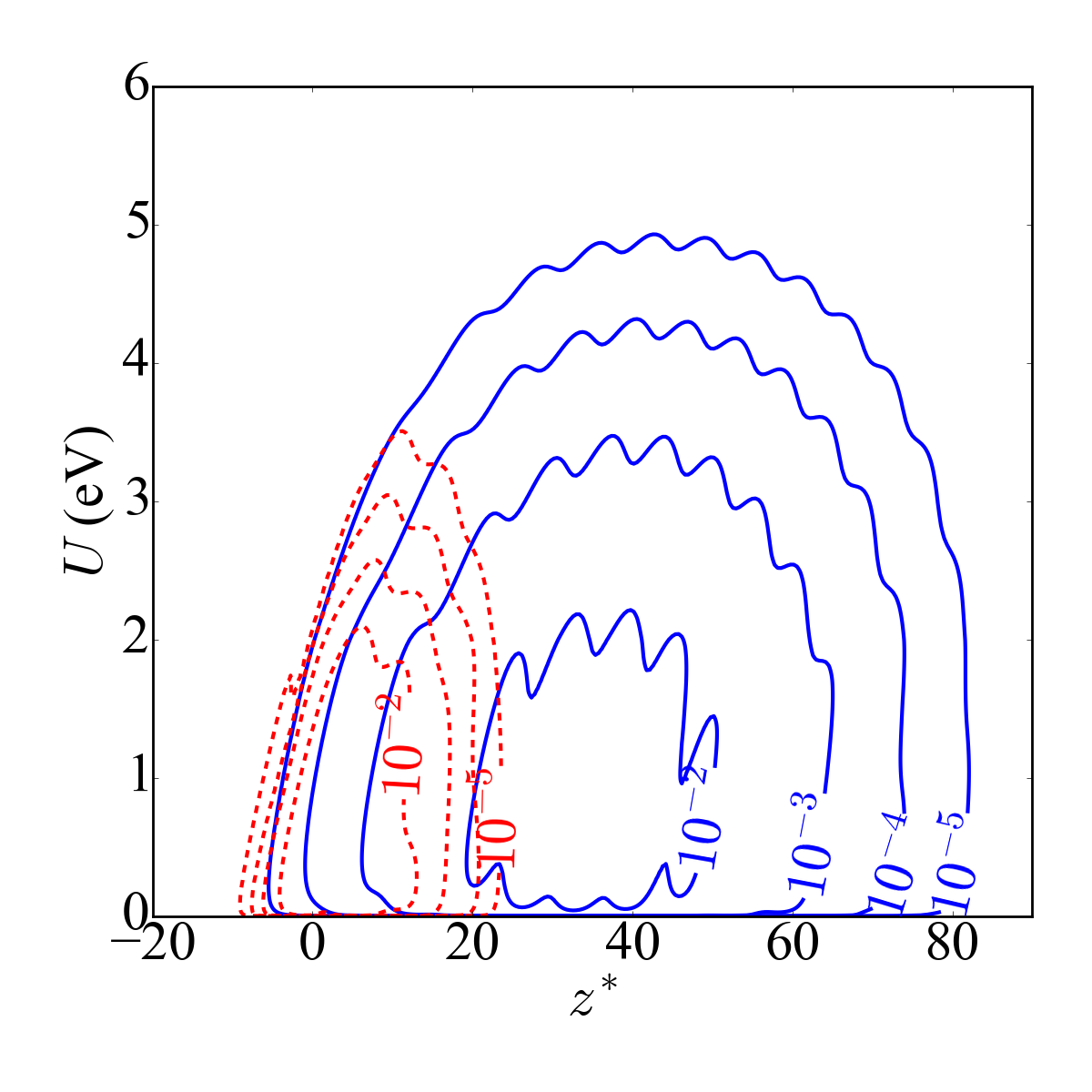}}%
\end{minipage}

\begin{minipage}[t]{1\columnwidth}%
\subfloat{\includegraphics[width=0.25\textwidth]{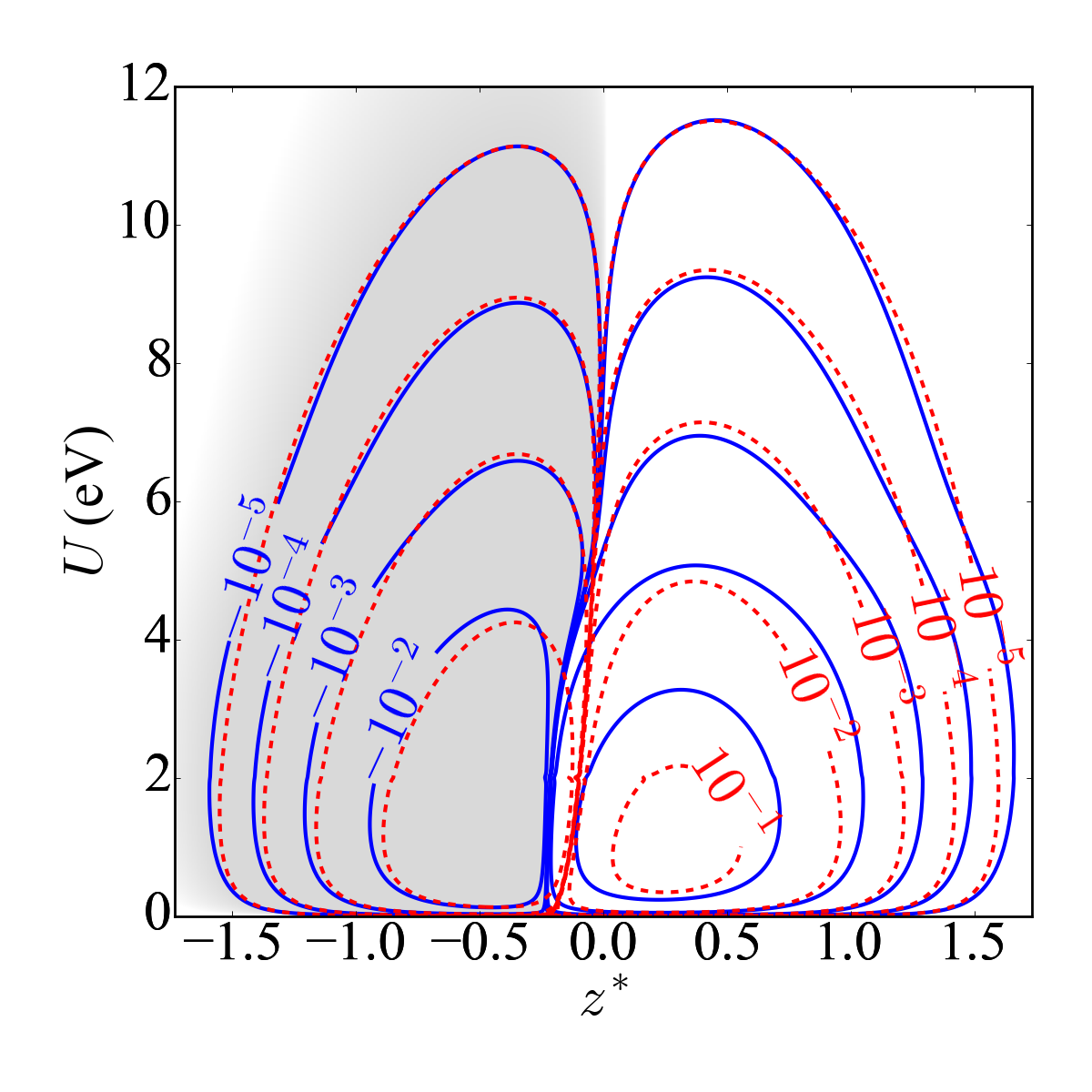}}\subfloat{\includegraphics[width=0.25\textwidth]{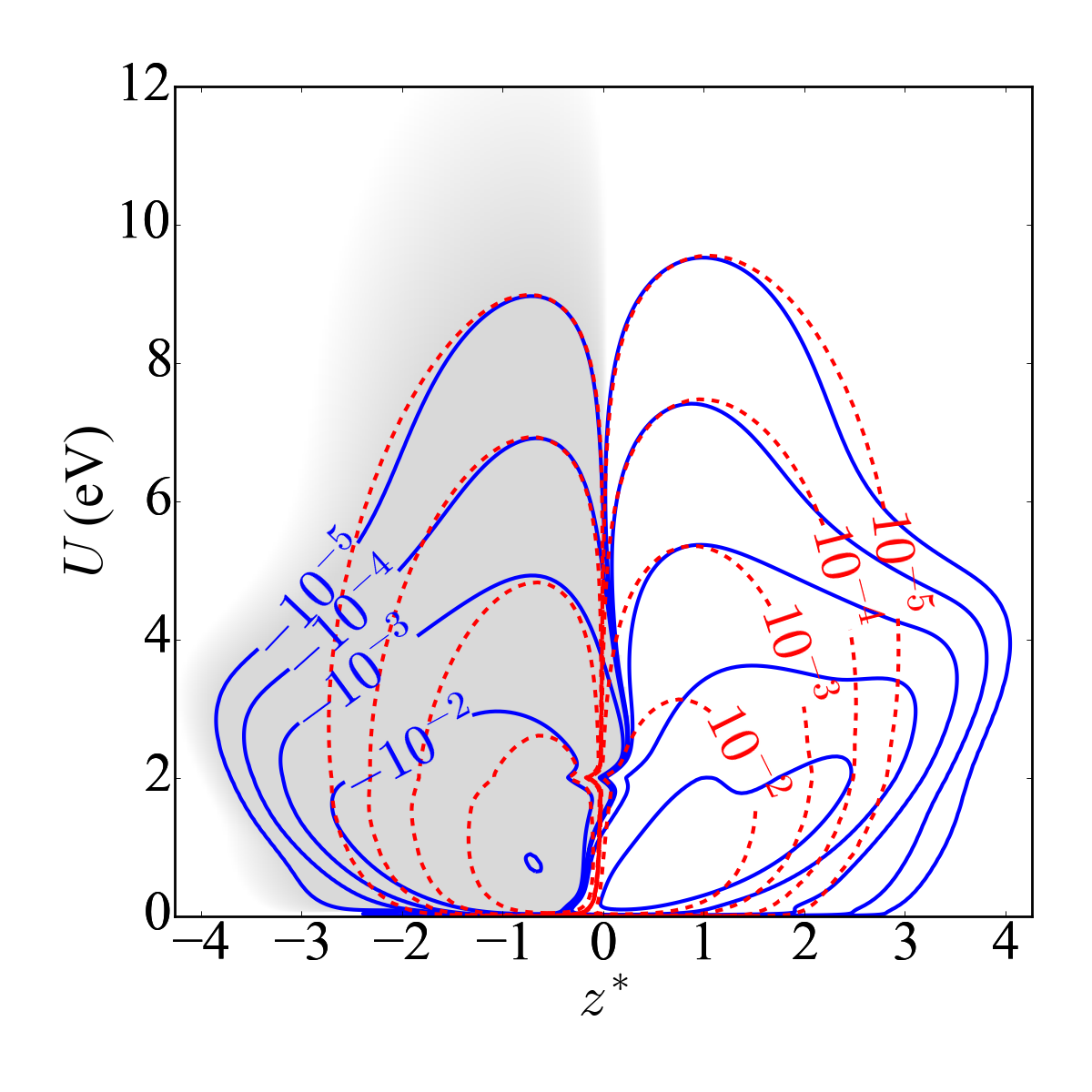}}\subfloat{\includegraphics[width=0.25\textwidth]{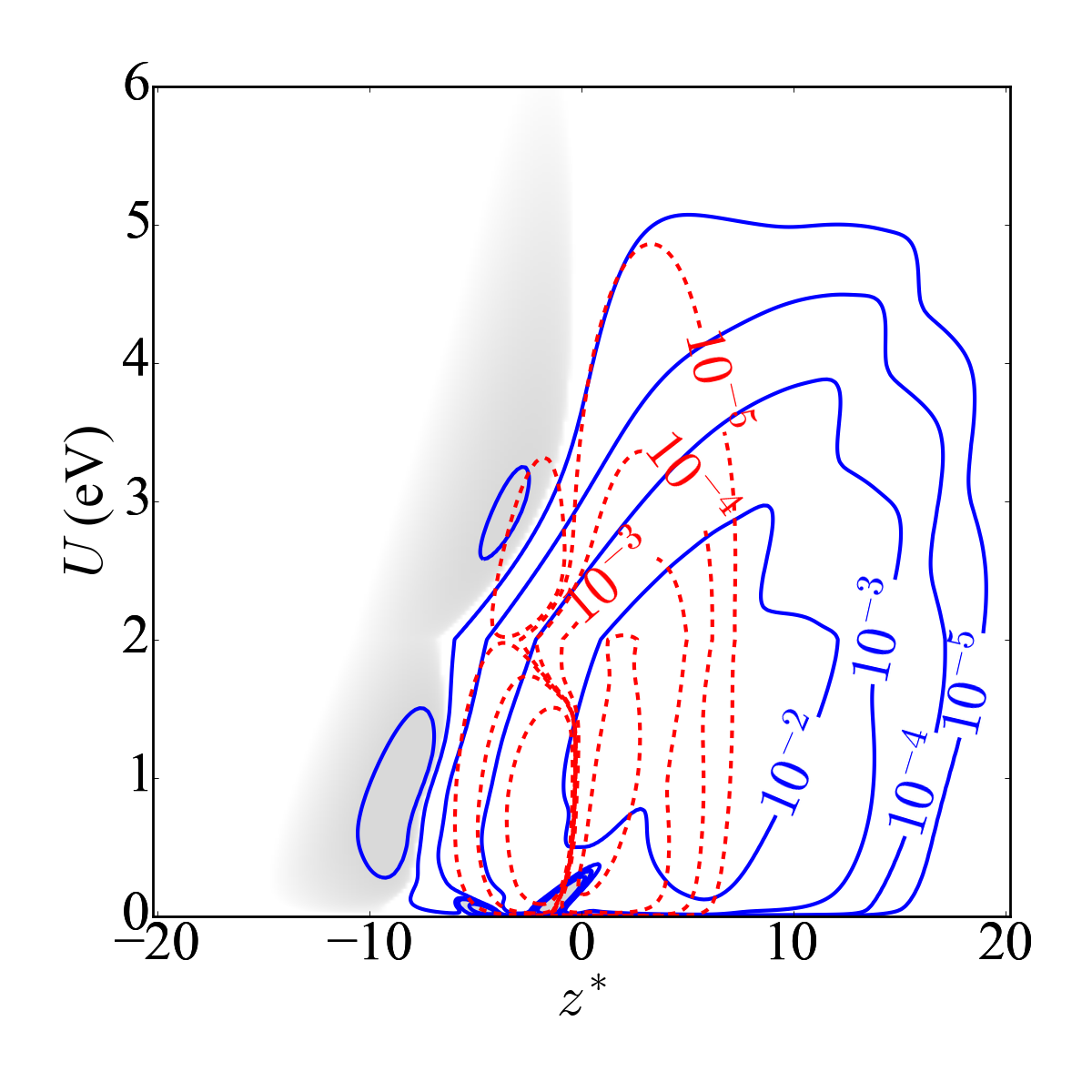}}\subfloat{\includegraphics[width=0.25\textwidth]{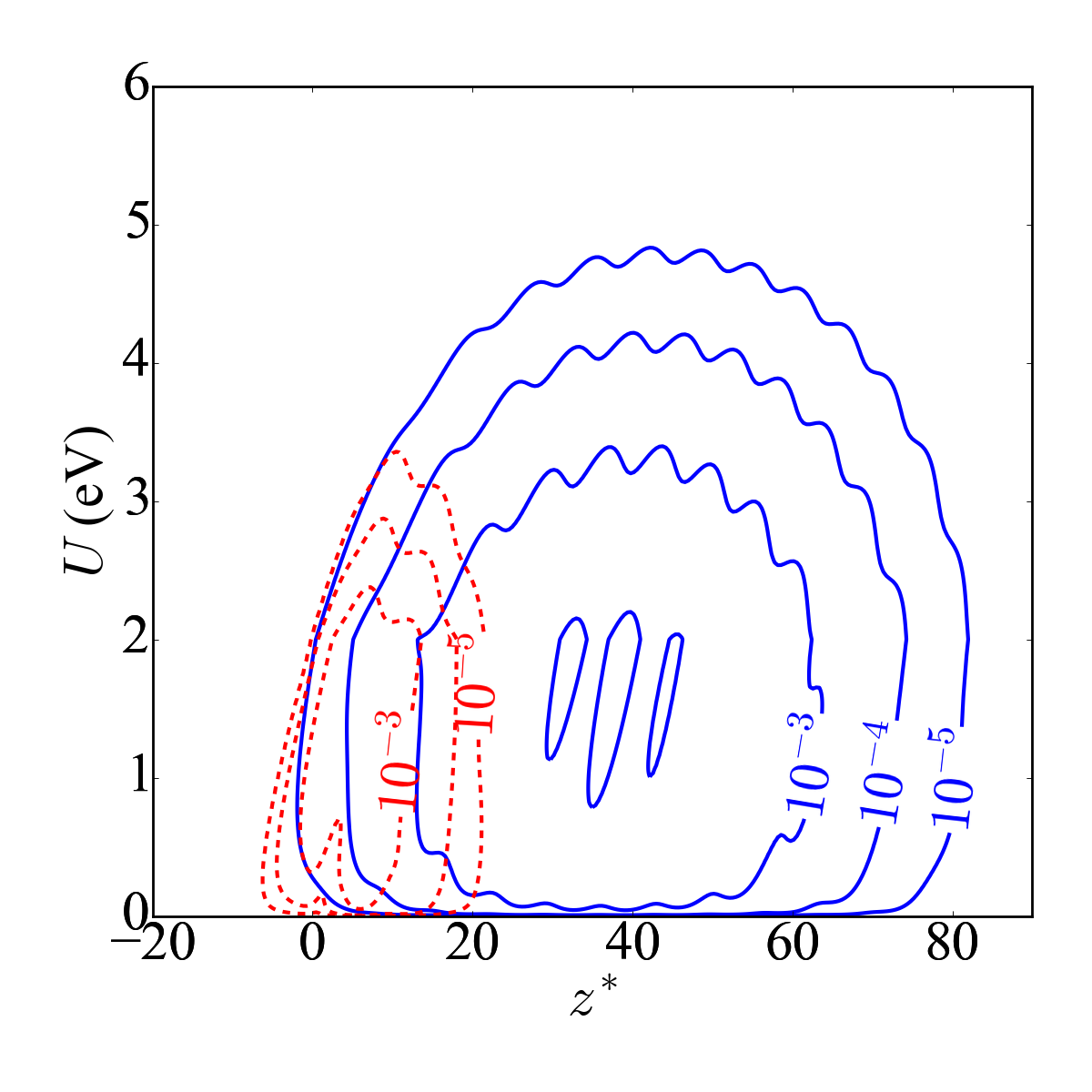}}%
\end{minipage}

\protect\caption{Temporal evolution of the distribution function components for model
(\ref{eq:HSmodel}) with $\Phi=0$ (dashed lines) and $\Phi=0.4$
(solid lines). The first row are $U^{1/2}f_{0}/n_{0}$ (eV$^{-1}$)
contours while the second row are $\left|Uf_{1}/n_{0}\right|$(eV$^{-1/2}$)
contours. The shaded contours indicate $Uf_{1}<0$. The four columns
represent the times, $t^{*}=0.2,\ 2,\ 20,$ and $200$ respectively.
\label{fig:ContourPhi}}
\end{sidewaysfigure}

\begin{sidewaysfigure}
\begin{minipage}[t]{1\columnwidth}%
\subfloat{\includegraphics[width=0.25\textwidth]{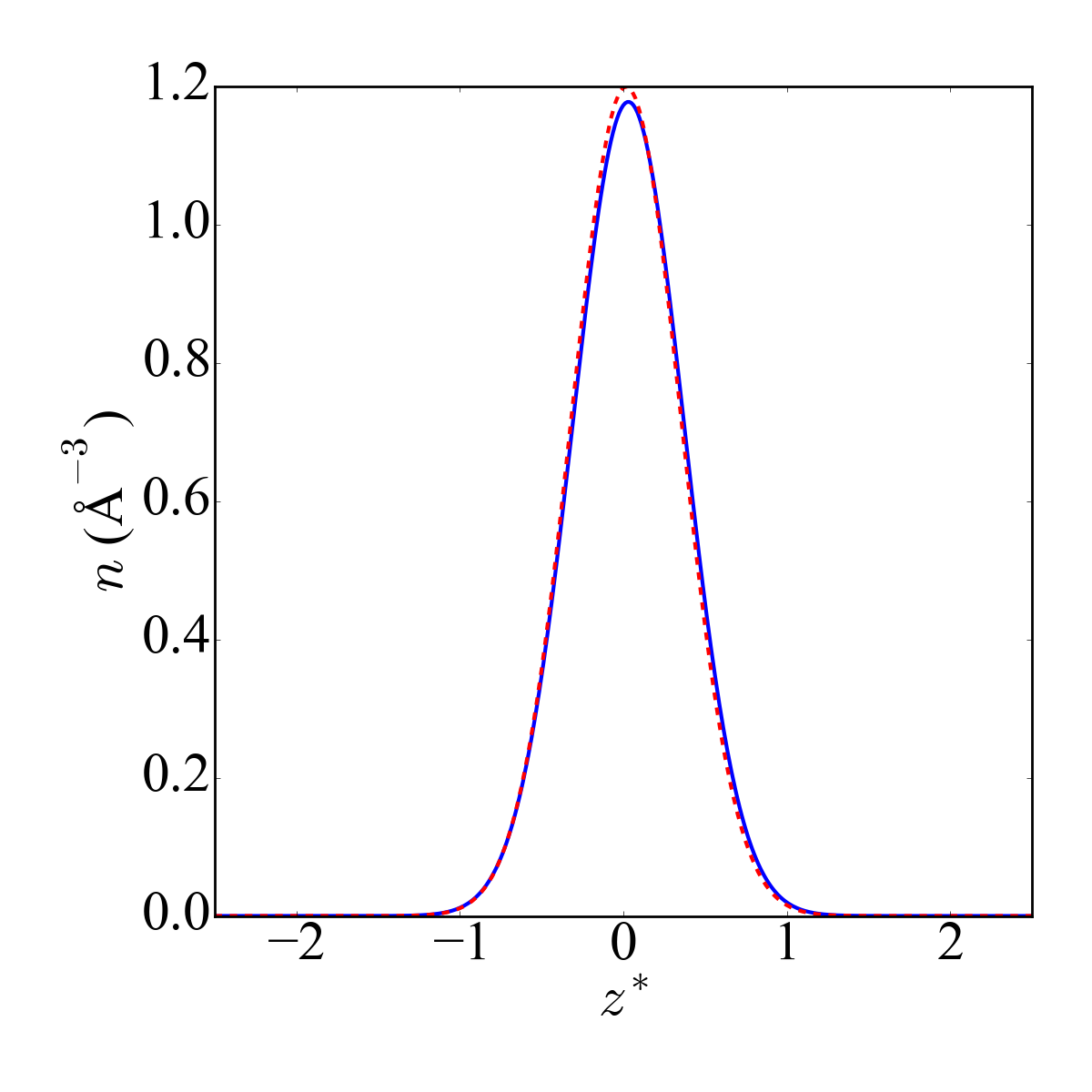}}\subfloat{\includegraphics[width=0.25\textwidth]{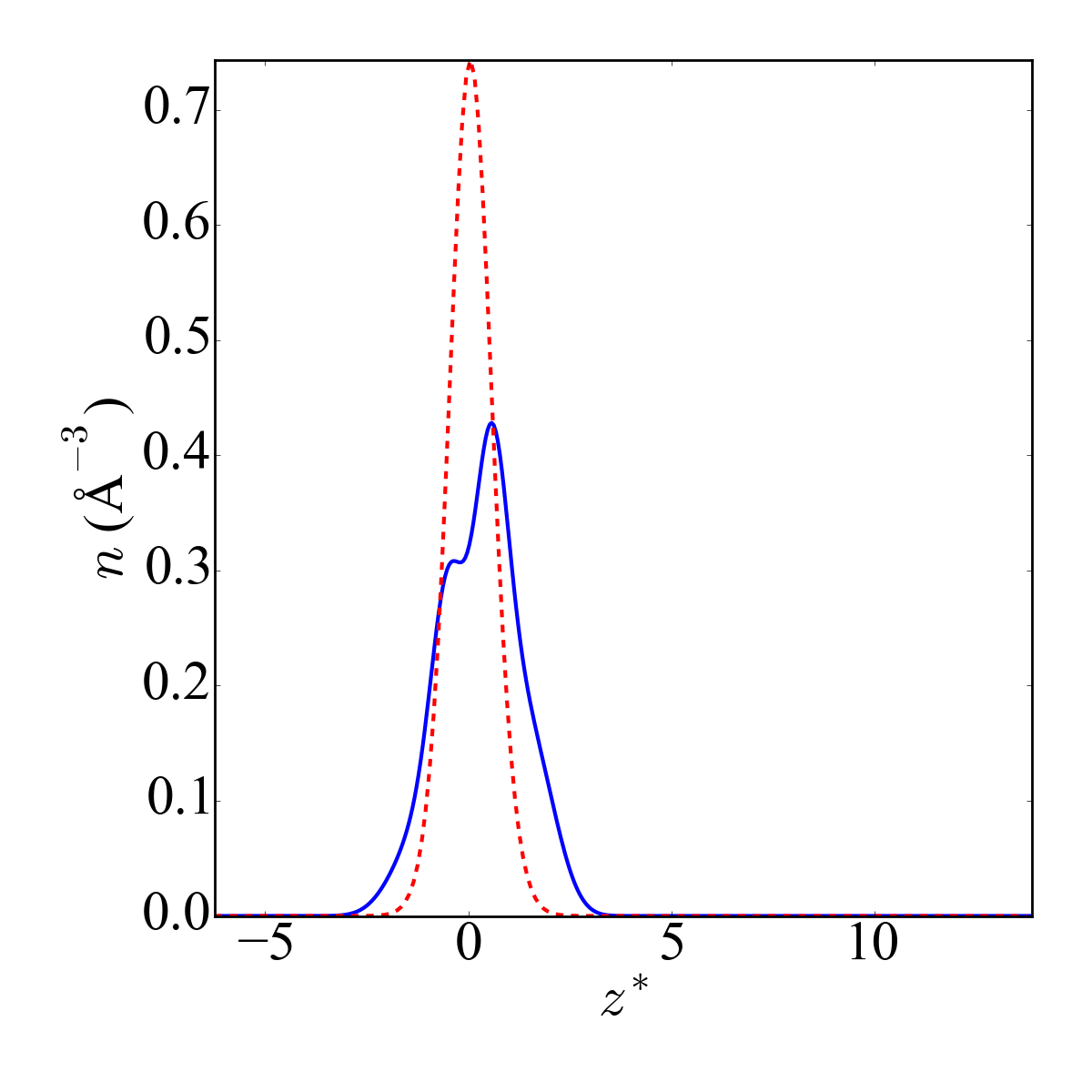}}\subfloat{\includegraphics[width=0.25\textwidth]{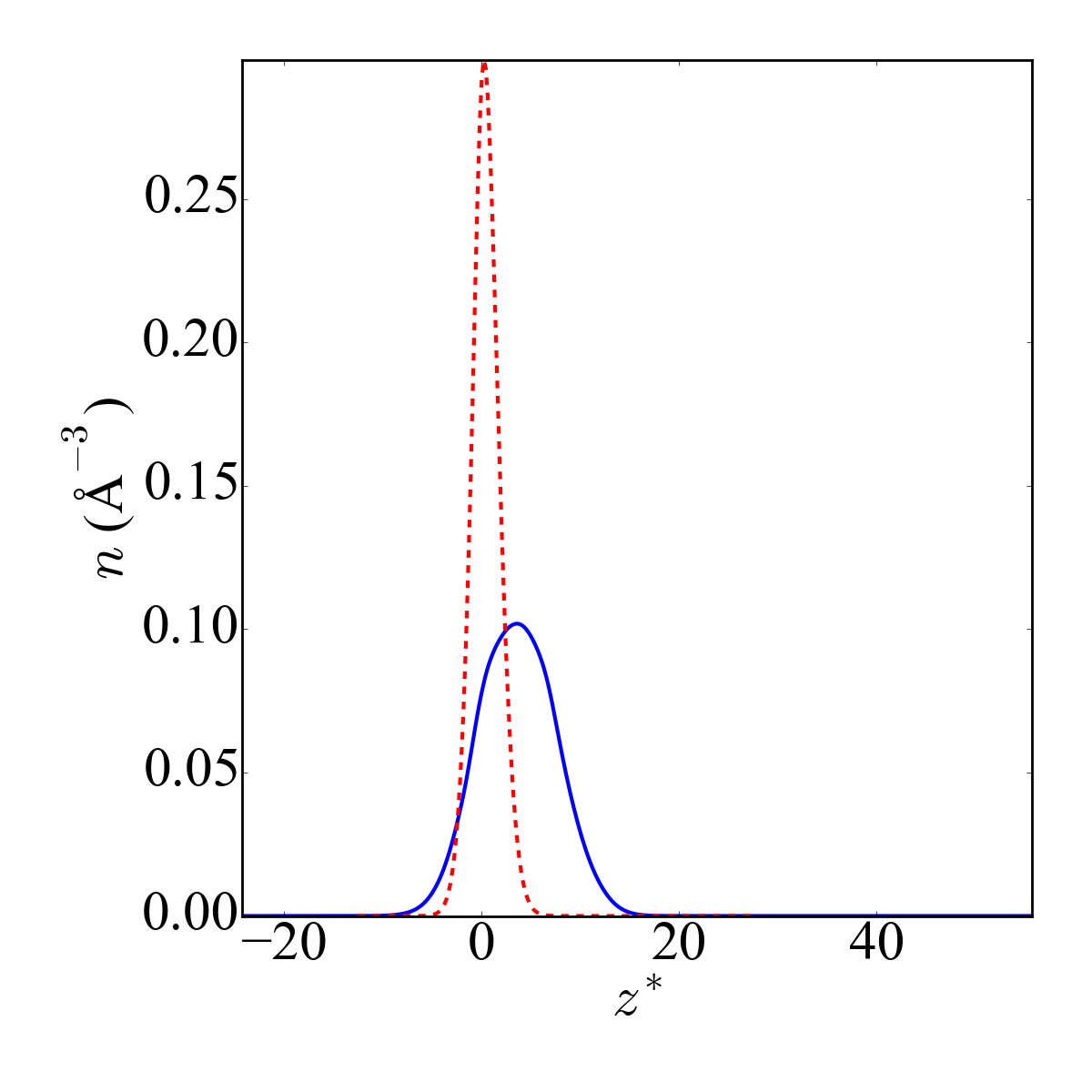}}\subfloat{\includegraphics[width=0.25\textwidth]{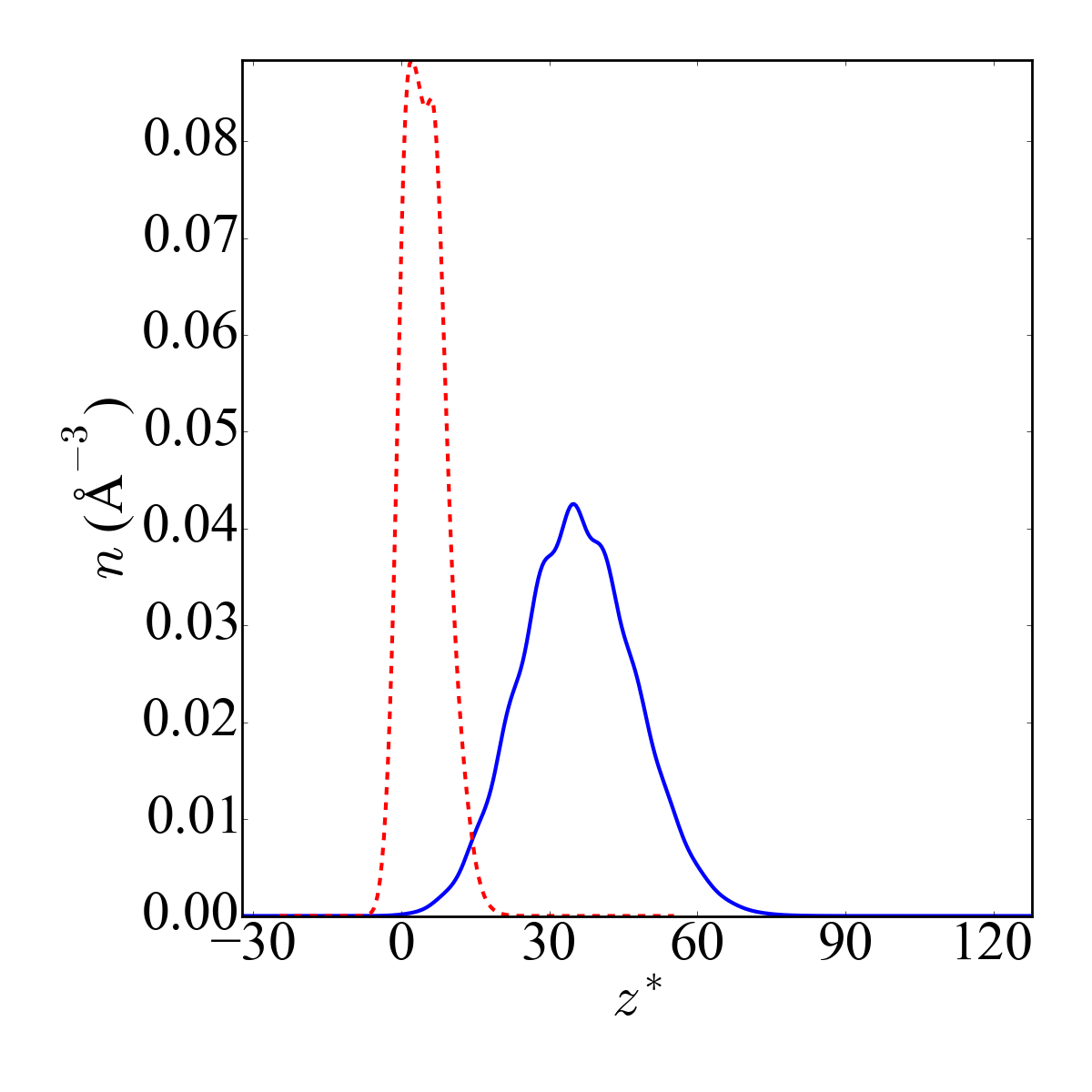}}%
\end{minipage}

\begin{minipage}[t]{1\columnwidth}%
\subfloat{\includegraphics[width=0.25\textwidth]{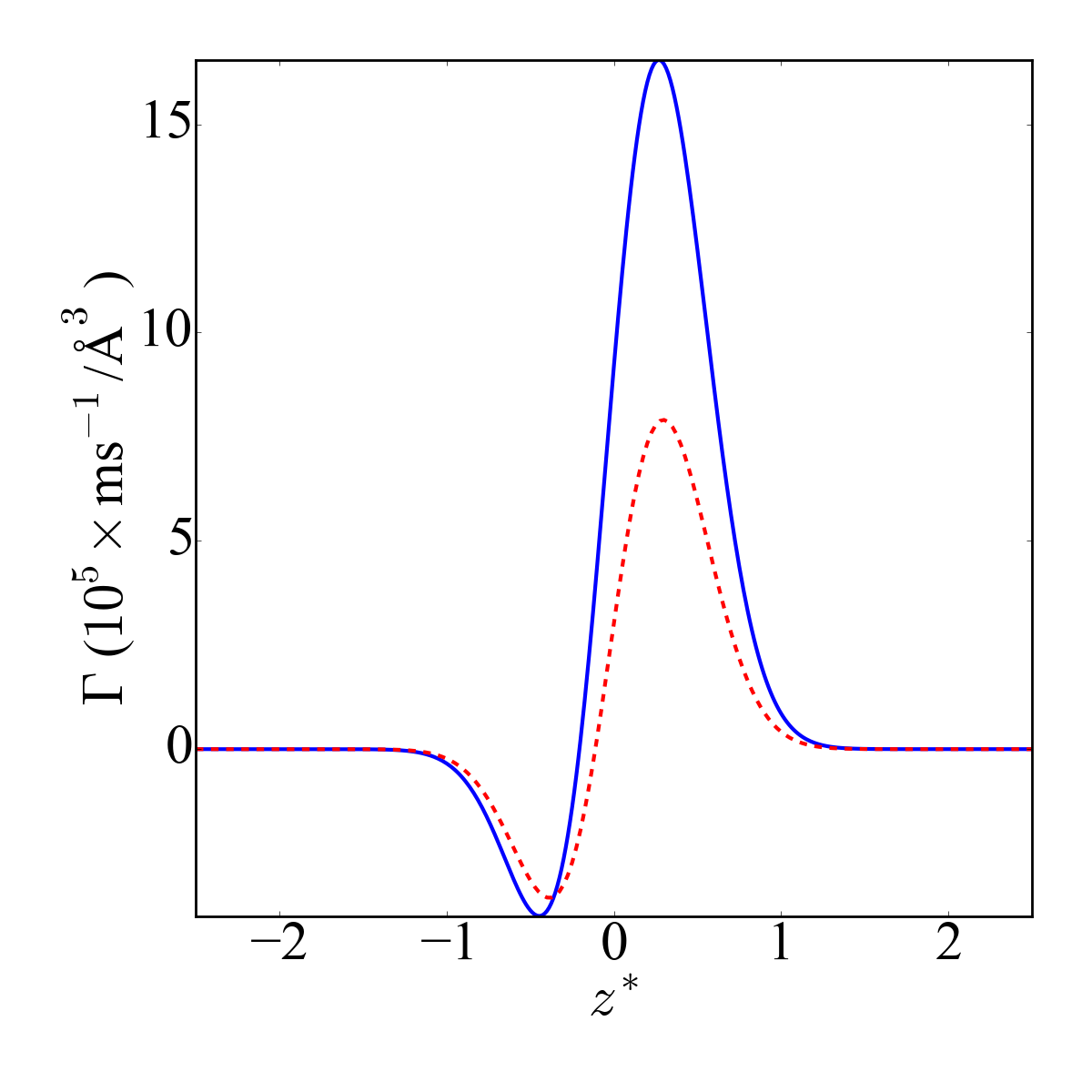}}\subfloat{\includegraphics[width=0.25\textwidth]{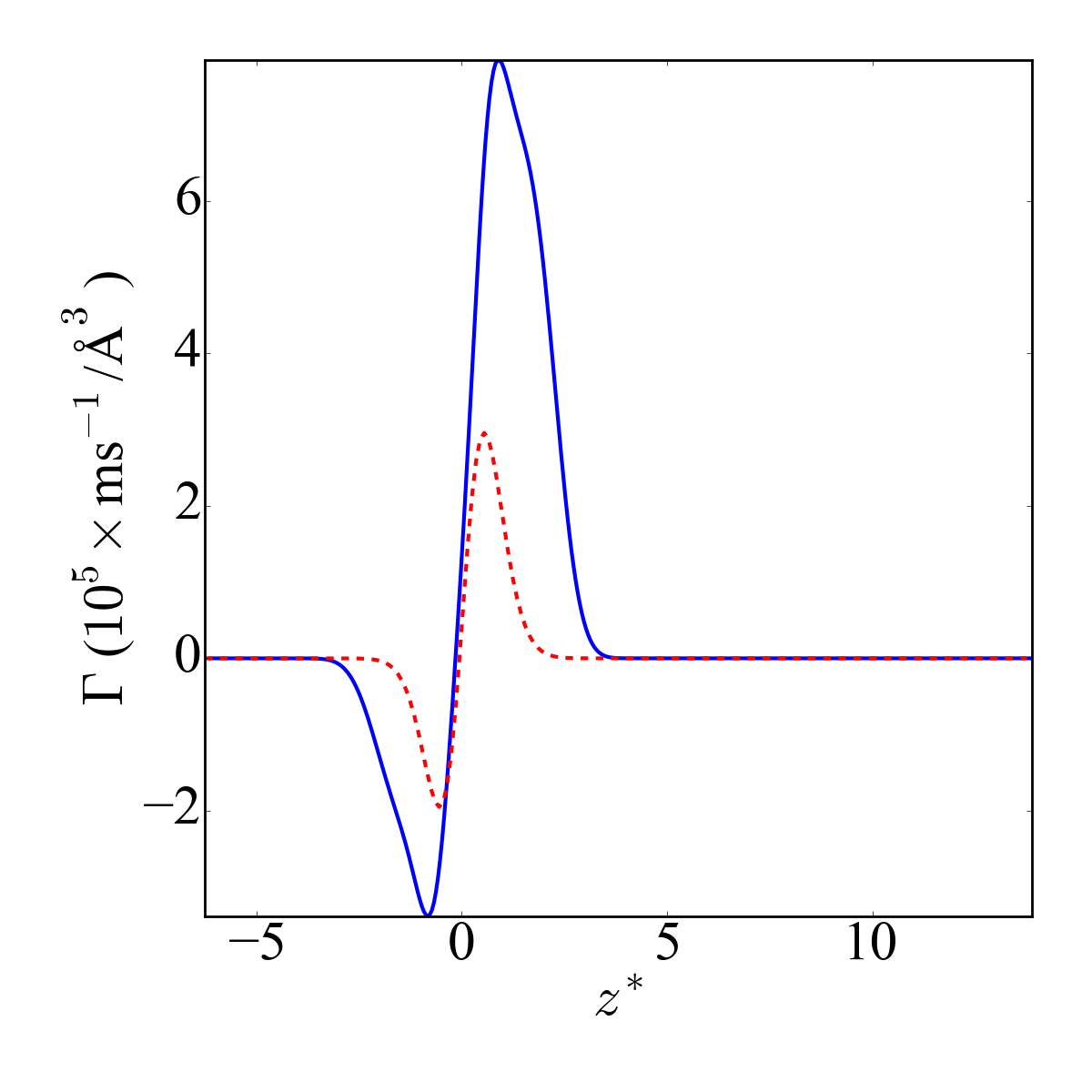}}\subfloat{\includegraphics[width=0.25\textwidth]{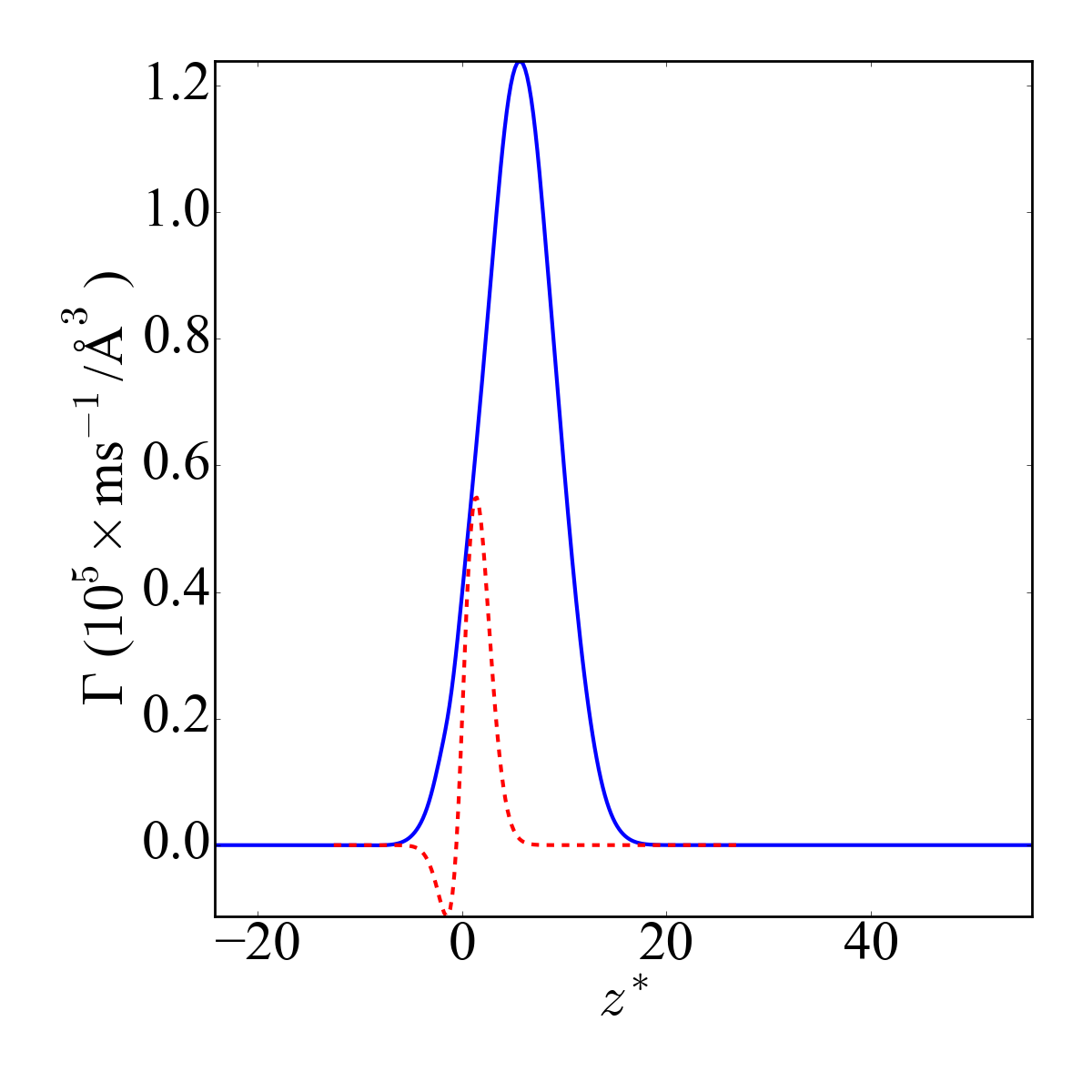}}\subfloat{\includegraphics[width=0.25\textwidth]{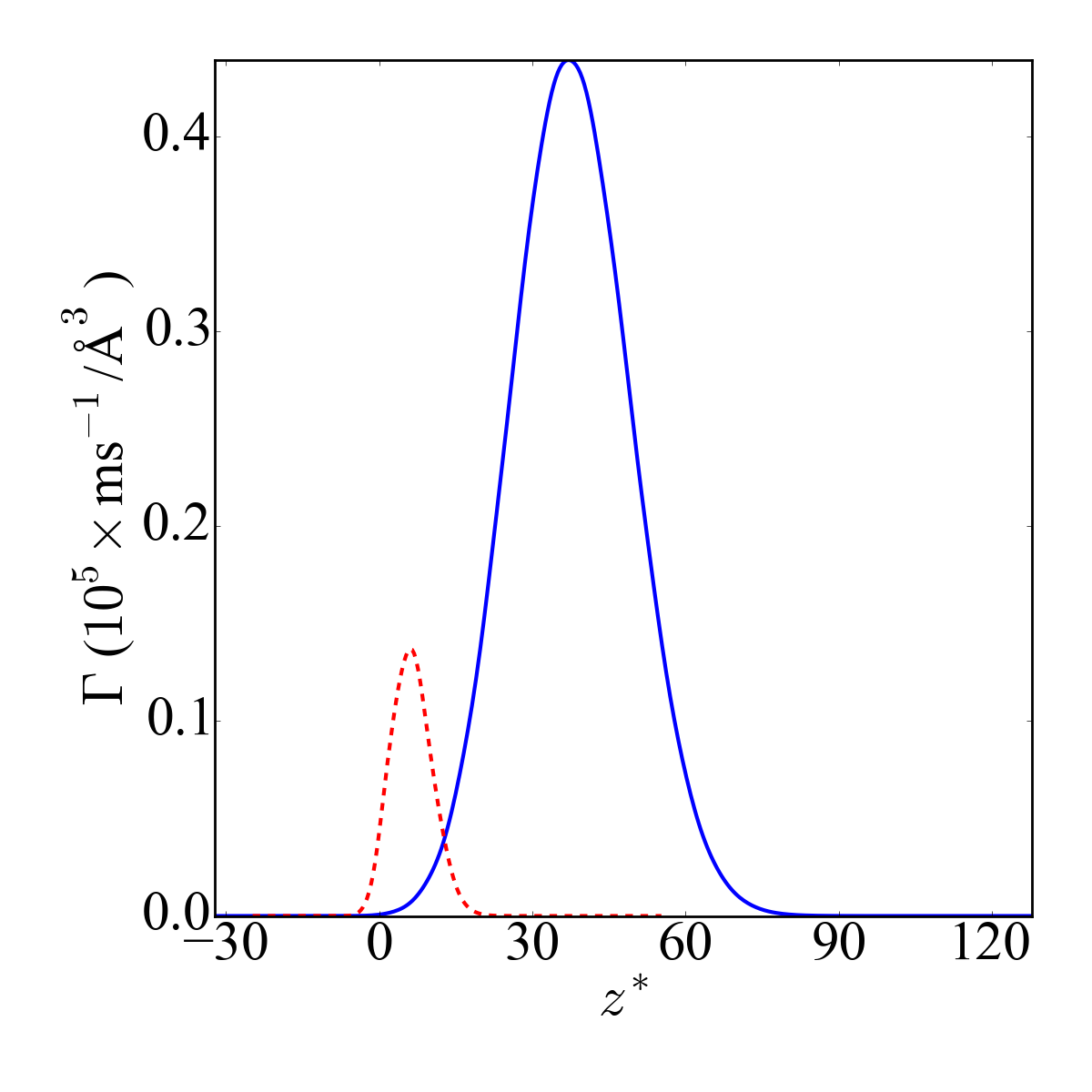}}%
\end{minipage}

\protect\caption{\label{fig:PTvel}Temporal evolution of the spatially varying density
and flux for model (\ref{eq:HSmodel}) with $\Phi=0$ (dashed lines)
and $\Phi=0.4$ (solid lines). The four columns represent the times,
$t^{*}=0.2,\ 2,\ 20,$ and $200$ respectively.}
\end{sidewaysfigure}

In Figure \ref{fig:ContourPhi} the space-time evolution of the $f_{0}$
and $f_{1}$ velocity distribution function components are compared
for $\Phi=0$ and $\Phi=0.4$ at three different times. The space-time
evolution of the integral moments of $f_{0}$, electron density $n(z,t)$,
and velocity moment of $f_{1}$, the flux $\Gamma$, are displayed
in Figure \ref{fig:PTvel}. The timescale for variation of $f_{0}$
is governed by $\sim\left(2\frac{m}{m_{0}}\nu_{m}\right)^{-1}$, and
hence there is no explicit $\Phi$ dependence in the timescale, however
differences arise due to the implicit energy dependence in the collision
frequency (which does depend on $\Phi)$ and the coupling to higher
order moments with different timescales. The timescale for variation
of $f_{1}$ on the other hand is governed by $\tilde{\nu}_{m}^{-1}$,
which has an explicit $\Phi$ dependence. The timescale for momentum
exchange is significantly decreased for increasing $\Phi$ at low
energies, as shown in Figure \ref{fig:PY-1}, however they approach
the same value at higher energies. We will show that this is reflected
in the evolution of the profiles. 

At small times (e.g. $t^{*}=0.2$), there are only small differences
in the $f_{0}$ contours between the two volume fractions, and this
is also highlighted in the density $n(z,t)$. At higher energies (>~5-6
eV) there are also very little differences in the $f_{1}$ contours
(reflecting the similarity in the momentum relaxation times at these
energies), however at low energies, the $\Phi=0.4$ contours for $f_{1}$
are significantly displaced in both energy and configuration-space
relative to the $\Phi=0$ case. This indicates significantly higher
advective and diffusive fluxes in this energy regime at this time,
which is evidenced in the flux profiles of Figure \ref{fig:PTvel}.
Given the sharp initial pulse with large spatial gradients, we observe
large positive and negative diffusive fluxes, along with a large positive
advective contribution. 

At larger times, the $f_{0}$ and $f_{1}$ contours in the $\Phi=0.4$
case depart significantly from the $\Phi=0$ case, initially in the
low energy regime and then finally the entire energy regime as the
higher energy electrons relax from the initial condition. The peaks
in each of the distribution components at larger times for $\Phi=0.4$
case are significantly displaced in the $z$-direction from the $\Phi=0$
case. This is reflected in both the density and flux profiles at larger
times, which highlight the enhanced drift and diffusion due to the
reduced momentum transfer cross-section associated with coherent scattering
for this model and field. Interestingly, at sufficiently long times,
the $\Phi=0.4$ contours have predominantly positive values, and only
very small negative excursions at low energies, in contrast to the
$\Phi=0$ contours. At these times, the flux is positive over the
swarm indicating that the advective contribution dominates the diffusion
contribution, since the density gradients are much more rapidly dissipated
in the $\Phi=0.4$ case, see Figure \ref{fig:PTvel}.

Strikingly, both the $\Phi=0$ and $\Phi=0.4$ contours for both $f_{0}$
and $f_{1}$ demonstrate periodic structures in both configuration
space and in energy space at sufficiently long times and sufficiently
downstream from the source. The periodic structures manifest themselves
earlier for the $\Phi=0.4$ case. These are the well known Franck-Hertz
oscillations \cite{FranHert14,Robsetal00}. A simplistic picture of
this non-hydrodynamic phenomena is that the electrons in the swarm
are being repeatedly accelerated by the electric field to an energy
above the inelastic process threshold whereby they undergo an inelastic
collision losing their energy. This simple physics is evidenced in
the $f_{0}$ and $f_{1}$ distributions. By integrating over the energy
to obtain the density and flux, shown in Figure~\ref{fig:PTvel},
much of the periodic structures observed in the distribution function
is masked, however some non-Gaussian spatial structure is still observed.
We will explore the $\Phi$-dependence of the wavelengths of oscillations
further in Section \ref{sub:Steady-state-Townsend-configurat}. 

\subsection{Steady-state Townsend configuration\label{sub:Steady-state-Townsend-configurat}}

\begin{figure}[H]
\protect\caption{Spatial variation of the average energy and average velocity under
SST conditions for model (\ref{eq:HSmodel}) with various volume fractions
$\Phi$.\label{fig:SST}}

\begin{minipage}[t]{1\columnwidth}%
\subfloat{\includegraphics[width=0.5\textwidth]{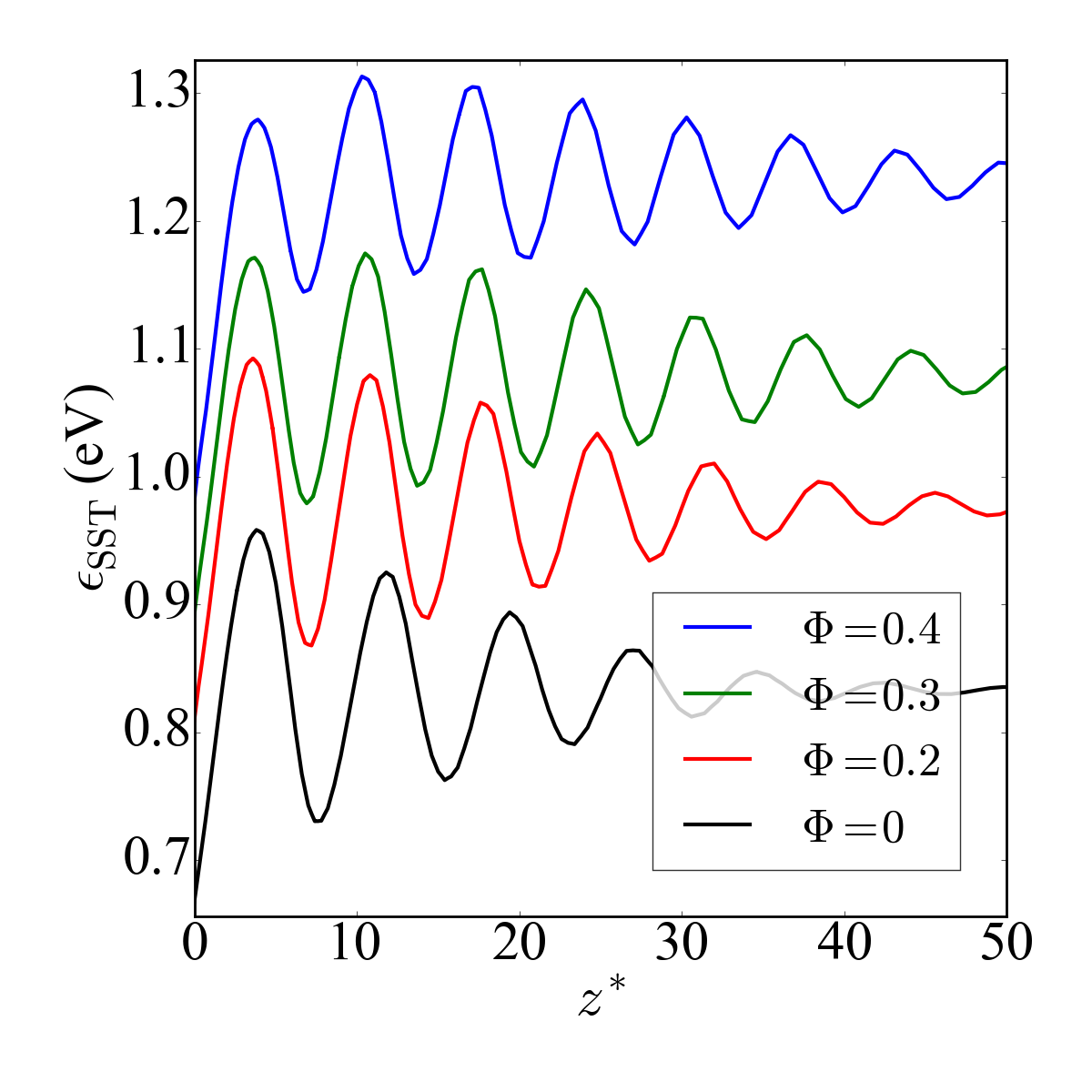}}~\subfloat{\includegraphics[width=0.5\textwidth]{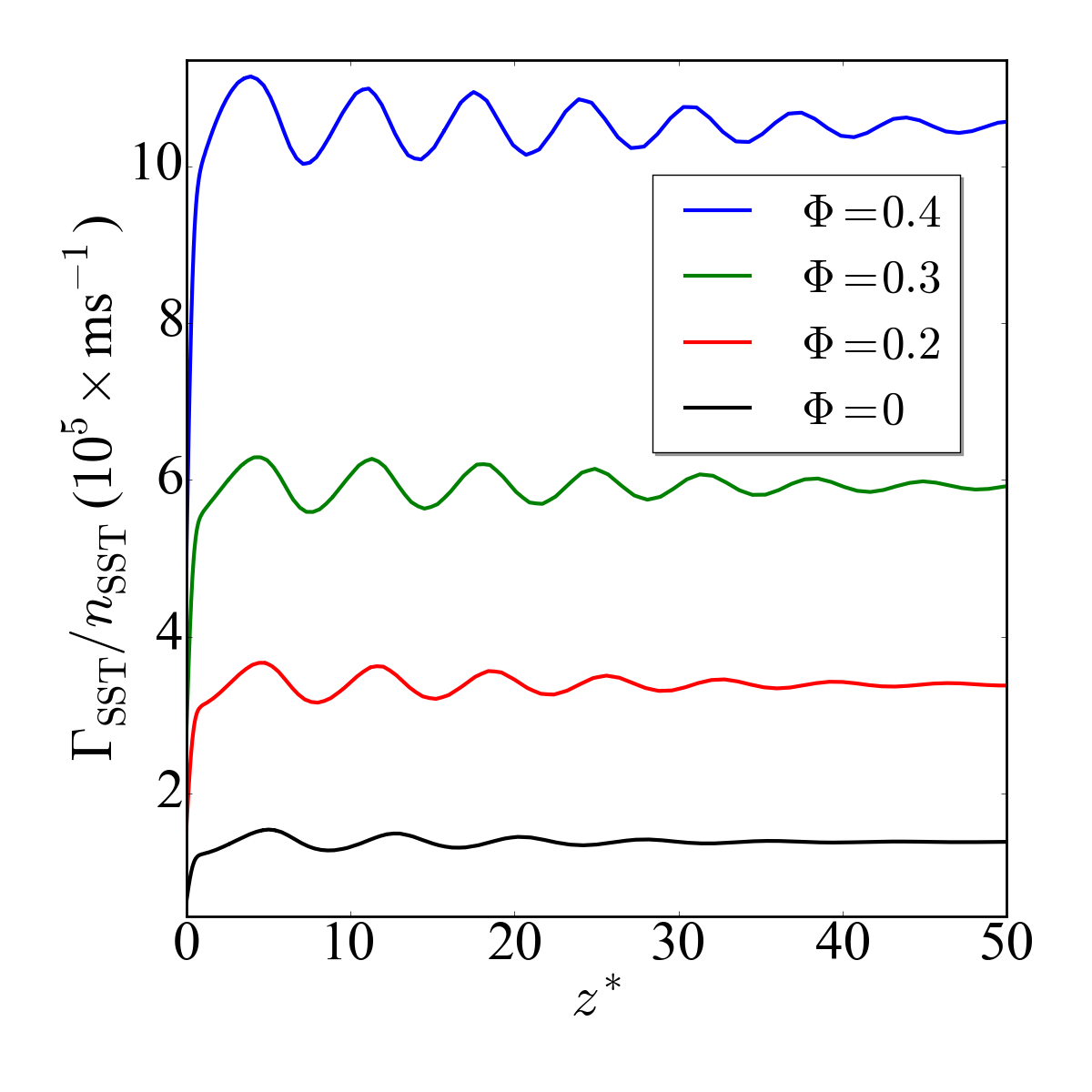}}%
\end{minipage}
\end{figure}

The solution detailed in Section \ref{sub:Space-time-evolutionPY}
is essentially equivalent to solving for the Boltzmann equation Green's
function for the model (\ref{eq:HSmodel}). A strict validation of
this approach and associated numerical code is to be able to reproduce
the Steady-State Townsend (SST) transport properties from the Green's
function solution, as described in Section \ref{sub:Green's-function-solution}.
The average energy (\ref{eq:energySST}) and average velocity (\ref{eq:fluxSST})
for SST simulations of various volume fractions are shown in Figure
\ref{fig:SST}. In the spatially asymptotic regime, the average energy
and the average velocity are equal to the hydrodynamic and pulsed-Townsend
values shown in Table \ref{tab:TransCompare}. It can be seen that
the SST properties demonstrate damped spatially periodic structures
similar to those observed in the Frank-Hertz experiment and other
investigations \cite{FranHert14,Flet85,Seguetal95,SigeWink97,PetrWink97,Robsetal00}.
They are a manifestation of the energy and space periodic structures
in the distribution function components, and in the spatially periodic
structures in the density and flux profiles of Figure \ref{fig:PTvel}.
By assuming the elastic scattering is weak, the width between the
peaks in the transport property profiles, $\lambda$, is directly
related to the threshold energy of the inelastic process, $U_{I}$
in eV, via \cite{White2012b}
\begin{equation}
\lambda=\frac{U_{I}}{\left(0.1\right)_{eV/Td}\left(E/n_{0}\right)_{\mbox{Td}}},\label{eq:FHwavelength}
\end{equation}
where the reduced electric field is in Townsend (Td). For model (\ref{eq:HSmodel}),
the theoretical spacing is $6.\dot{6}$. In Figure \ref{fig:SST}
it is possible to see that there are variations in the wavelength
of the spatial structures with $\Phi$, as well as significant differences
in the decay rates of the oscillation amplitudes. For $\Phi=0,$ the
wavelength is approximately $8.24\pm0.02$ and this decreases to $6.67\pm0.02$
for $\Phi=0.4.$ The differences arise explicitly due the differences
in the elastic momentum transfer cross-section, as well as implicit
variations associated with the modification to the swarm's energy
with $\Phi$. For $\Phi=0.4$, the momentum transfer cross-section
for elastic scattering is significantly reduced compared to the $\Phi=0$
case. Hence, the randomizing collisions that dampen the oscillations
\cite{Robsetal00} are reduced for $\Phi=0.4$ as compared to other
$\Phi$, and the variation of damping with $\Phi$ then follows. Likewise,
it should not be surprising that the wavelength for the $\Phi=0.4$
case is closest to the analytic value of (\ref{eq:FHwavelength}),
since the reduced momentum transfer associated with the $\Phi=0.4$,
more closely approximates the weak elastic scattering assumption used
in deriving it. 

We must also point out that the validity of these profiles are dependent
on the discretization of the distributions in configuration-space.
If the spatial discretization is of the same order as the Frank-Hertz
wavelength, then it will be very difficult to resolve these features
in the distributions and consequently the time-averaged profiles.
Of course, our initial choice for the discretization is small enough
to easily resolve these features, but as the simulation progresses
and the distribution diffuses, our adaptive mesh will increase in
range and also increase the spatial discretization step size. After
a point, the coarseness of the discretization causes the distribution
to slowly lose its features, which is visible in the time-averaged
quantities by the suppression of the amplitude of the oscillations.
In our simulations we expect our results for $z^{*}\gtrapprox30$
deviate from the true spatially dependent steady-state values, however
the fully relaxed values agree with the hydrodynamic values. It is
simple to address this issue by increasing the number of points in
configuration-space but this is also significantly more computationally
intensive.

\section{Spatio-temporal relaxation of electrons in liquid argon\label{sec:Electrons-in-dilute}}

Electron transport in liquid argon is an essential component in the
function of Liquid Argon Time Projection Chambers (LArTPC) which are
currently being used for high energy particle detection \cite{Marc13}.
Ionized electrons in liquid argon originating from the high energy
particles are accelerated under the action of an electric field to
generate a current and consequently reconstruct the path of the high
energy particle. Typically these chambers operate with electric field
strengths of less than $500$ kV/cm. The aim of this component is
to follow the spatio-temporal evolution of these ionized electrons
in liquid argon, relevant to the operation of these detectors. Foxe
et al. \cite{Foxetal15} have measured the energy distribution of
the electrons ionized by high energy particles in liquid argon, and
have shown that the majority of the ionized electrons have energies
below $1$ eV. Consequently in this study we employ an initial source
energy-distribution that is constant in energy space up to $1$ eV,
i.e.,
\begin{equation}
f_{U}(U)=CU^{-1/2}\Theta\left(U-1\,\mathrm{eV}\right),\label{eq:sourceAr1}
\end{equation}
where $\Theta$ is the Heaviside step function, and $U$ is in eV
and $C$ is a normalisation constant. The mean energy of this distribution
is $0.5$ eV. The swarm is released from a narrow Gaussian in configuration-space,
\begin{equation}
f_{z}(z)=\frac{1}{\Delta z_{0}\sqrt{2\pi}}\exp\left(-\left(\frac{z}{\Delta z_{0}}\right)^{2}\right)\label{eq:sourceAr2}
\end{equation}
so that the full initial phase-space distribution is $f\left(U,z,0\right)=Af_{U}(U)f_{z}(z)$,
where $A$ is a normalization constant such that $\int U^{1/2}f\left(U,z,0\right)dU=1.$
For argon, we take $\Delta z_{0}=10$, a larger initial spread than
for the Percus-Yevick model, reflecting the smaller cross-sections
of argon.

\subsection{Cross-sections, potentials and screening}

\begin{figure}[H]
\protect\caption{\label{fig:Liquid-Ar-cross-sections} The momentum transfer cross-sections
in the gas-phase (dashed line) and liquid-phase (solid line) for electrons
in argon \cite{Boyletal15}.}

\centering{}\includegraphics[width=0.5\textwidth]{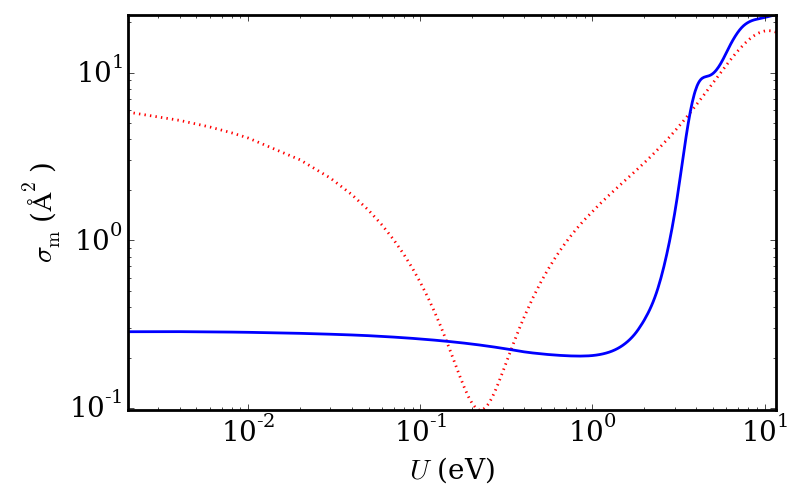}
\end{figure}

In a recent paper \cite{Boyletal15} we investigated the modifications
required to treat transport of electrons in dense gaseous and liquid
argon, with our simulations focused purely on the hydrodynamic regime.
The work followed closely the approach of Lekner and Cohen \cite{Lekn67,CoheLekn67}
updated using modern scattering theory techniques, theories, potentials
etc. The potentials, screening factor, and cross-sections derived
in \cite{Boyletal15} are used once again here, and we briefly summarize
the process. The gas-phase elastic scattering cross-sections were
calculated by solving the Dirac-Fock scattering equations \cite{Chen08},
and were shown to give good agreement with both beam scattering cross-sections
and swarm experimental drift velocities and characteristic energies.
As the medium's density increased, density effects became important.
The effects of the high density of the liquid are included in our
calculations through several modifications of the gas-phase scattering
properties, and are dependent on the liquid argon pair-correlation
function and associated static structure factor \cite{Yarnell73}.
The first modification is to account for the screening of a single
induced atomic dipole by the induced dipoles of all other atoms. In
a high density medium, the potential produced by the induced multipole
moments is of a sufficiently long enough range to interact with induced
multipole moments by other atoms in the bulk. The second modification
is to construct an effective potential in which there are contributions
from both the target atom and the surrounding bulk, the latter of
which is done as an ensemble average. The momentum transfer cross-sections
calculated from the dilute gaseous and liquid argon potentials are
shown in Figure \ref{fig:Liquid-Ar-cross-sections}. It is significant
to note the absence of the Ramsauer minimum in the liquid-phase cross-section.

\subsection{Results}

To consider conditions representative of those in liquid state particle
detectors, we simulate electron transport in liquid argon under the
following conditions:
\begin{eqnarray}
E/n_{0} & = & 2.5\times10^{-3}\ \mbox{Td,}\nonumber \\
T & = & 85\ \mbox{K,}\\
m_{0} & = & 40\ \mbox{amu}.\nonumber 
\end{eqnarray}
The reduced field is equivalent to $500$ kV/cm with a density corresponding
to liquid argon, $n_{0}=0.0213\ \mbox{\AA}^{-3}$. For this reduced
electric field and source distribution, given in (\ref{eq:sourceAr1})-(\ref{eq:sourceAr2}),
the electron swarm energies are generally well below the first inelastic
channel threshold energy ($8.9$ eV), so that there is no inelastic
channel operative, and hence the periodic spatial structures observed
in the Percus-Yevick hard-sphere liquid model above are not present. 

\begin{sidewaysfigure}
\begin{minipage}[t]{1\columnwidth}%
\subfloat{\includegraphics[width=0.25\textwidth]{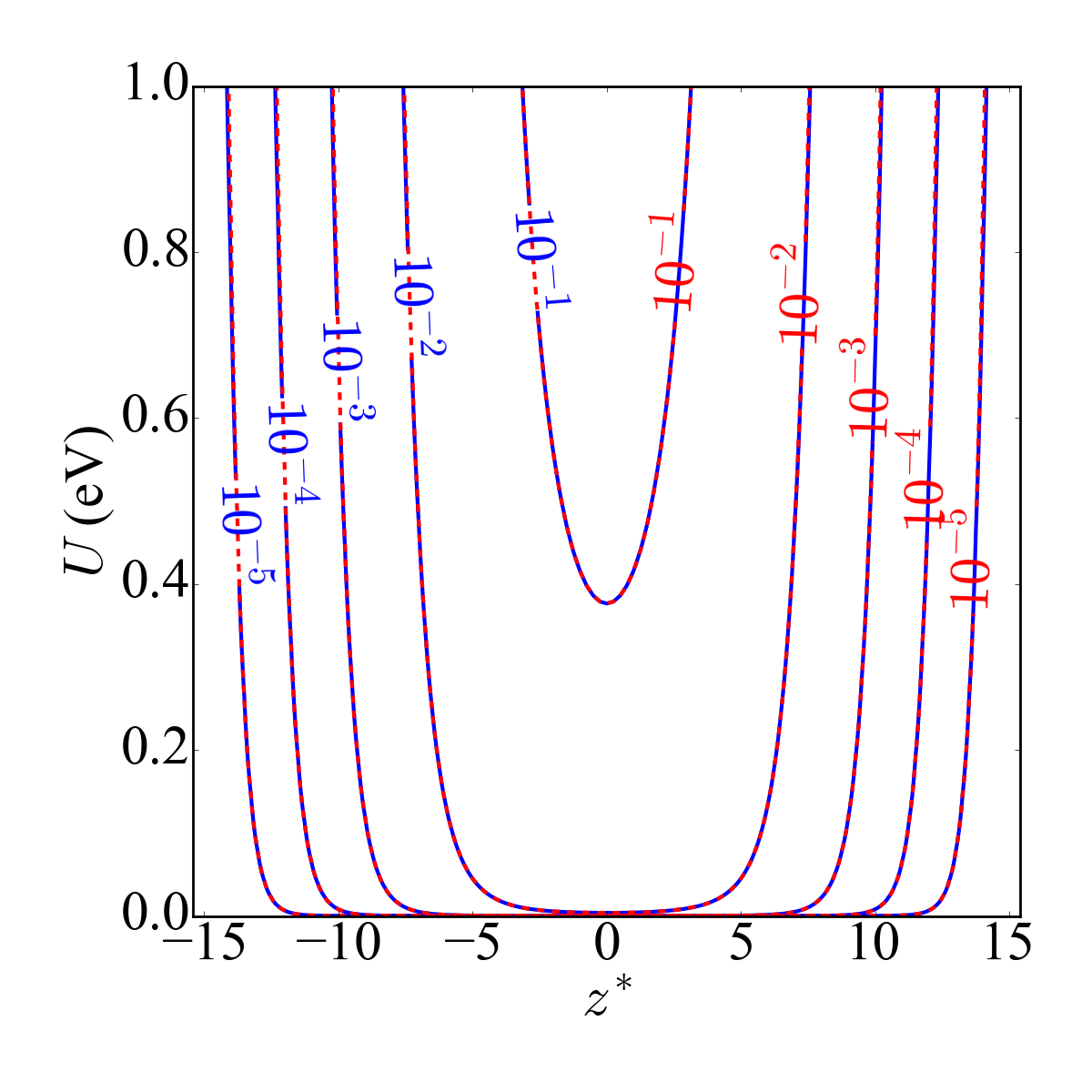}}\subfloat{\includegraphics[width=0.25\textwidth]{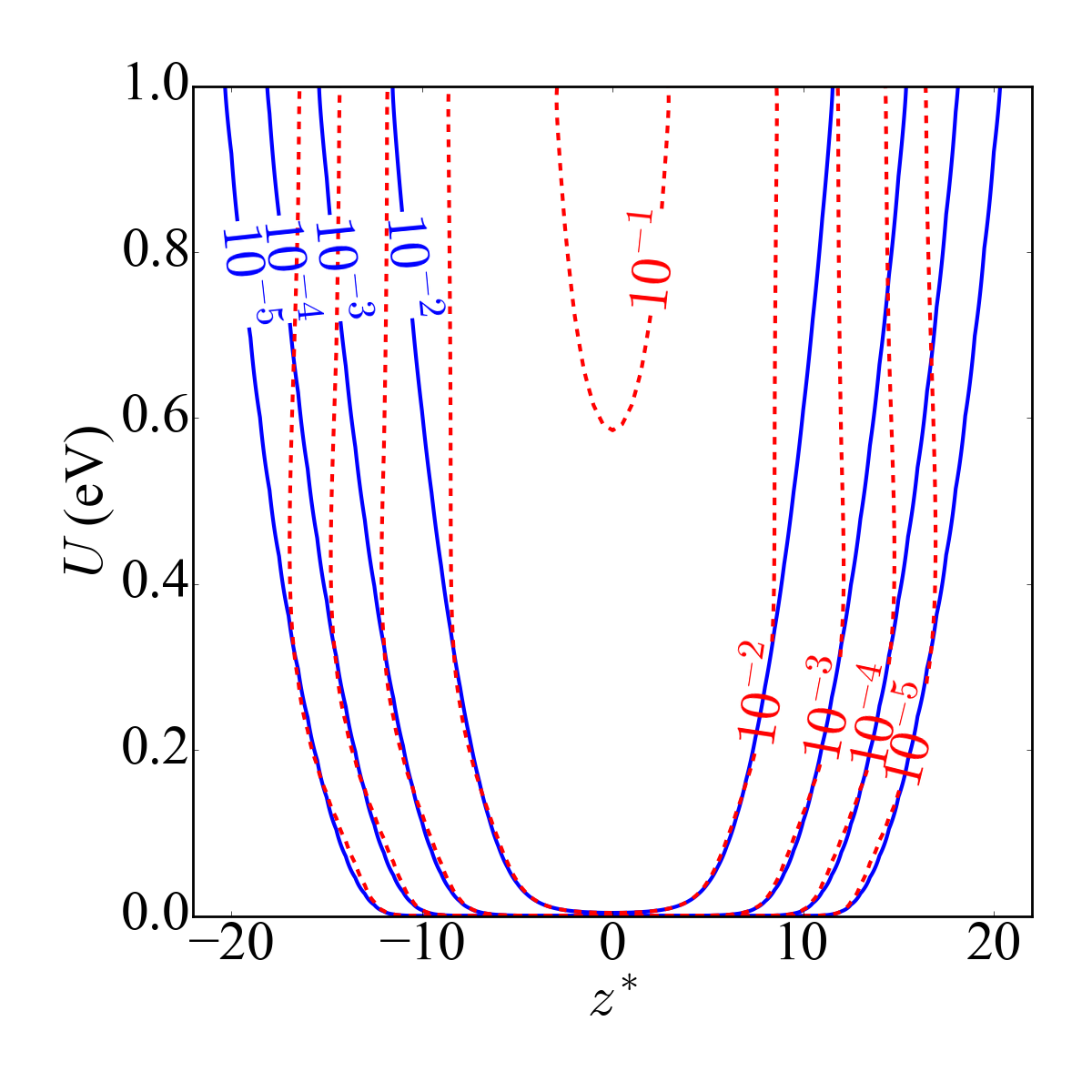}}\subfloat{\includegraphics[width=0.25\textwidth]{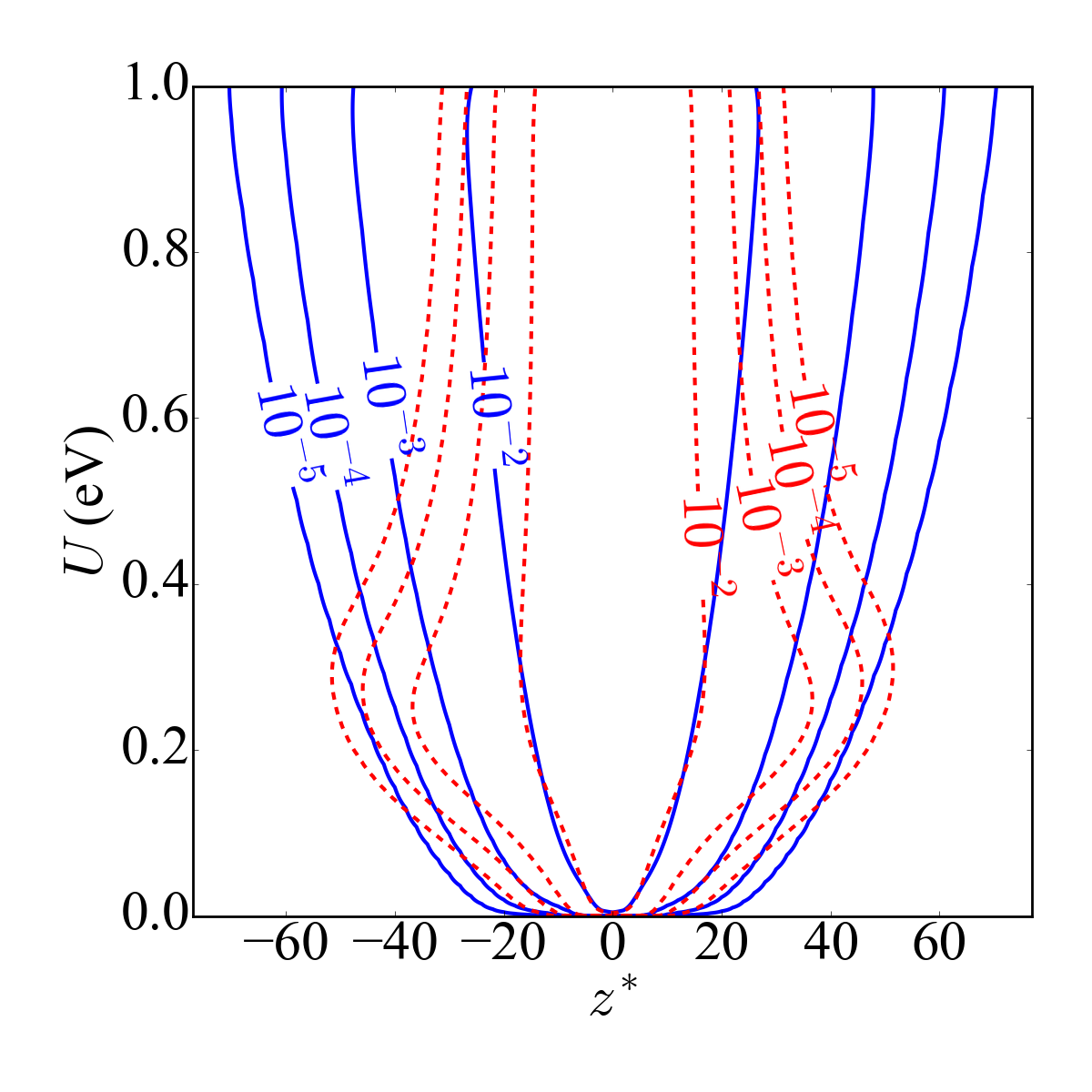}}%
\end{minipage}

\begin{minipage}[t]{1\columnwidth}%
\subfloat{\includegraphics[width=0.25\textwidth]{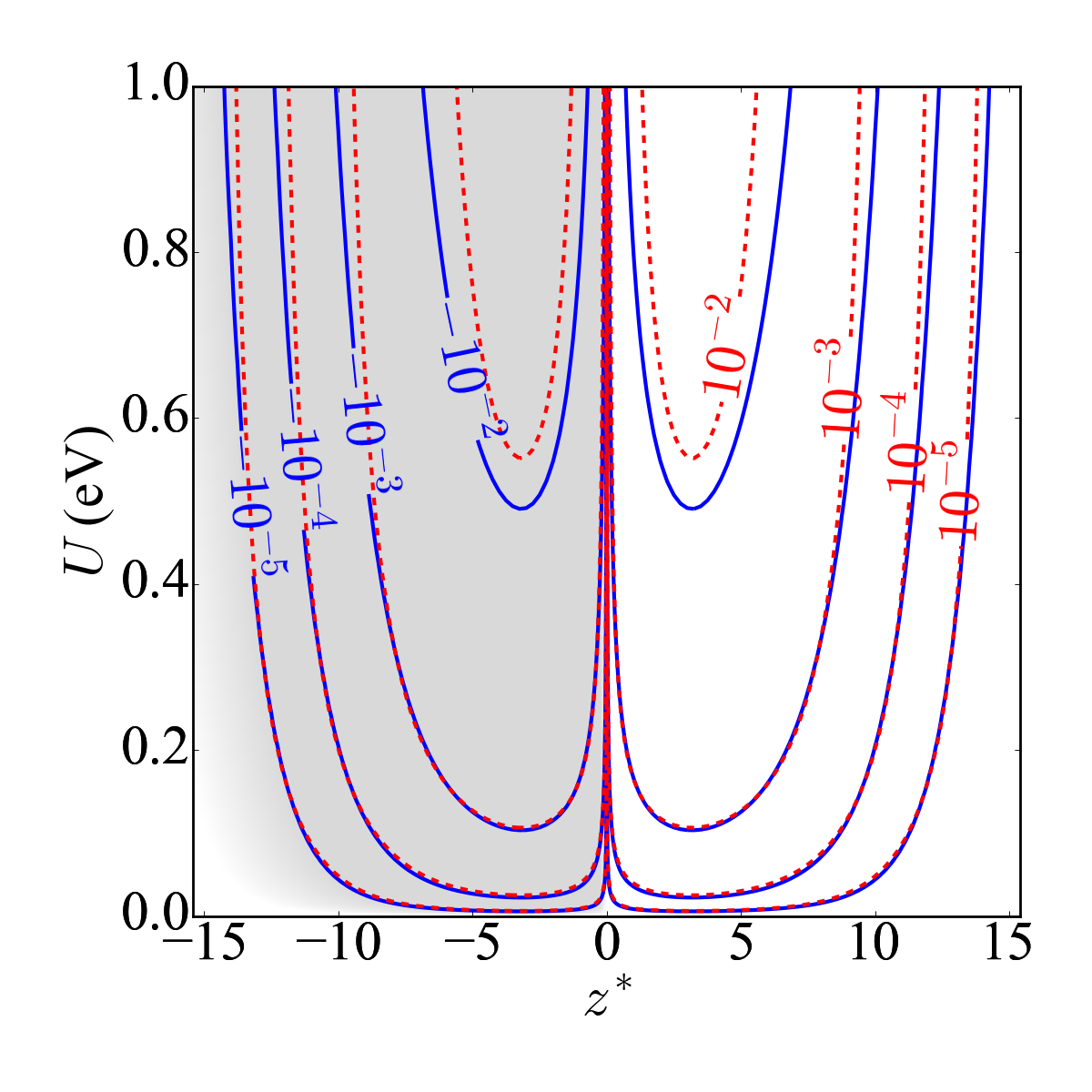}}\subfloat{\includegraphics[width=0.25\textwidth]{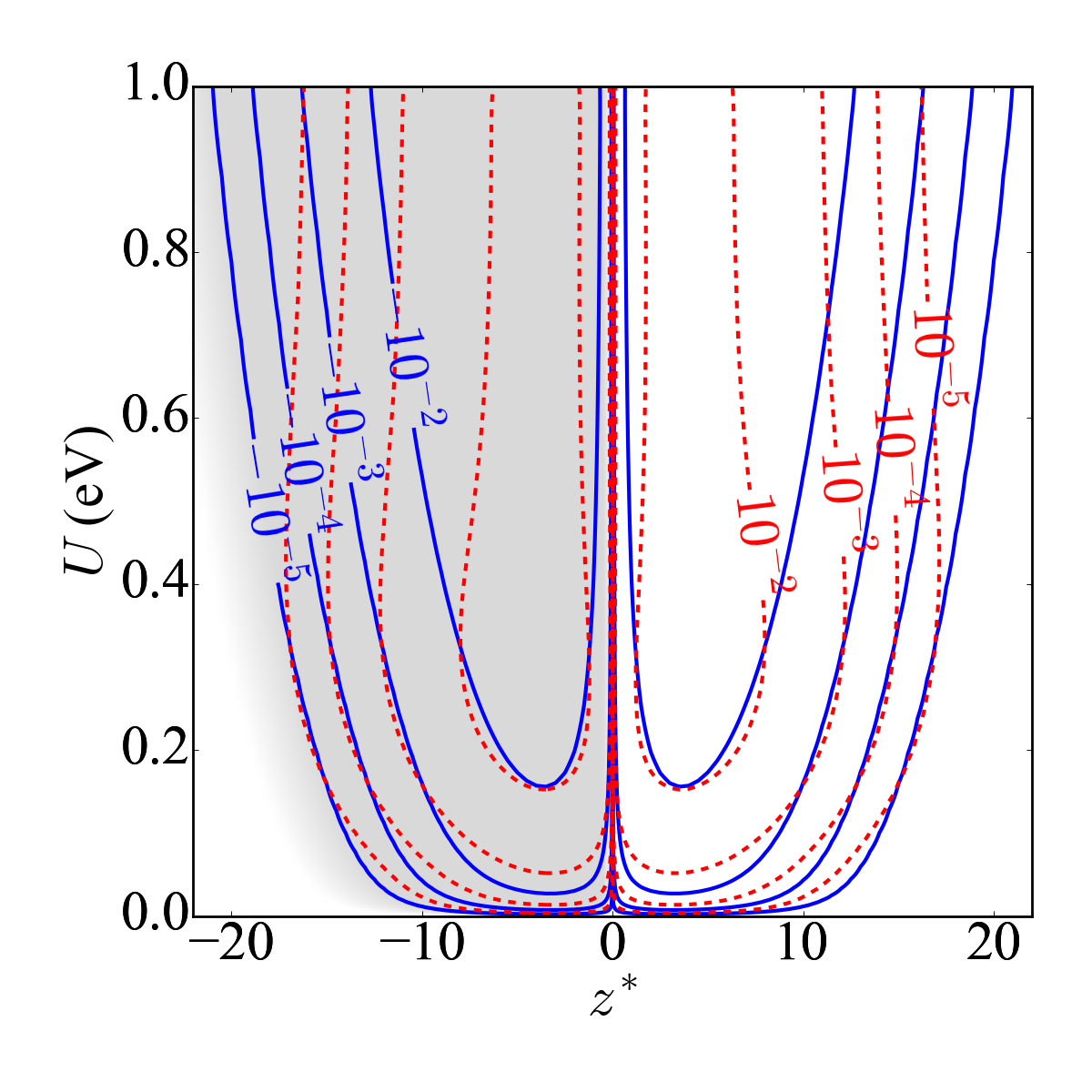}}\subfloat{\includegraphics[width=0.25\textwidth]{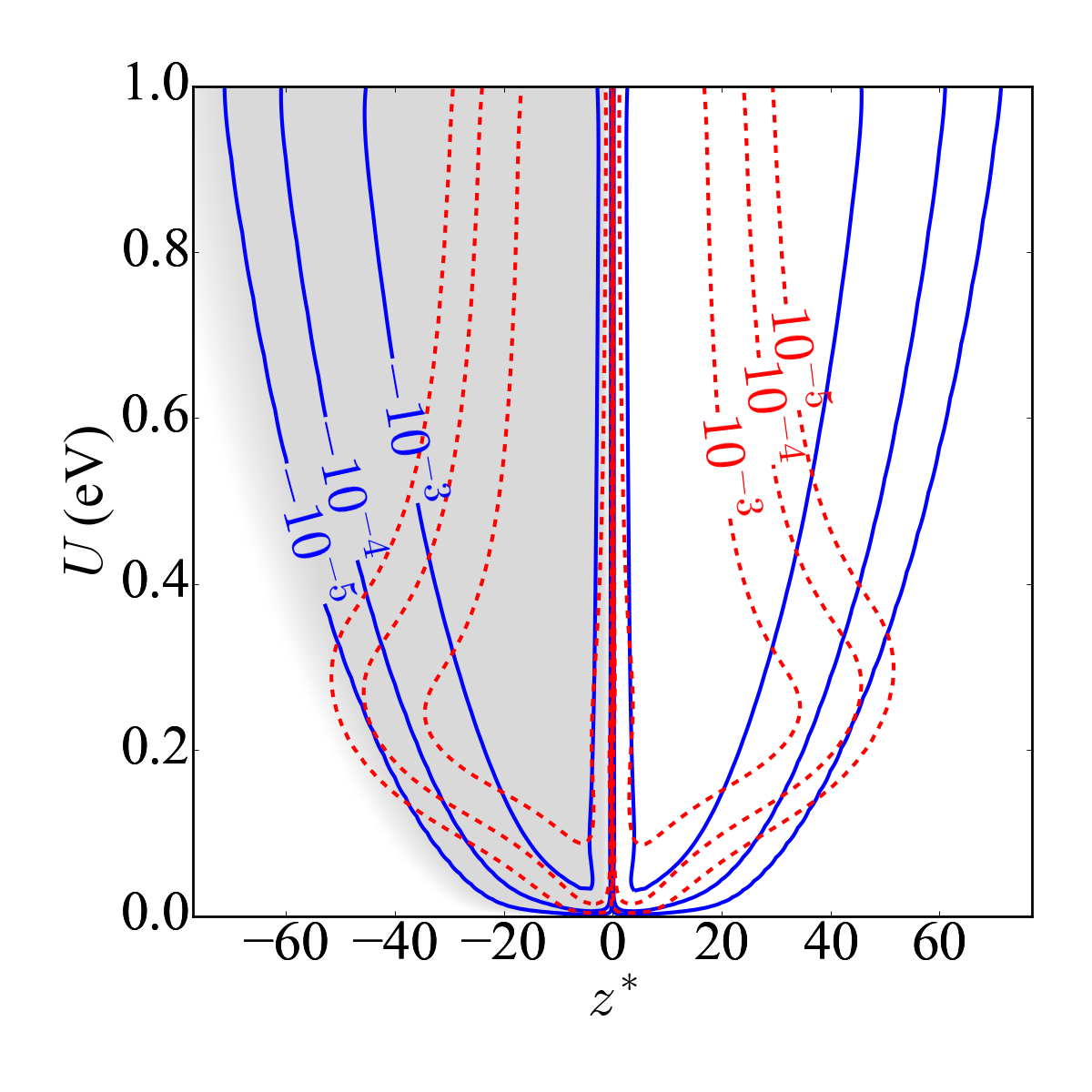}}%
\end{minipage}\protect\caption{\label{fig:ContourArg}Temporal evolution of the distribution function
components for gas-phase (dashed lines) and liquid-phase (solid lines)
argon. The first row are $U^{1/2}f_{0}/n_{0}$(eV$^{-1}$) contours,
the second row are $\left|Uf_{1}/n_{0}\right|$(eV$^{-1/2}$) contours.
The shaded contours indicate $Uf_{1}<0$. The three columns represent
the times, $t^{*}=1,\ 10,\mbox{ and }100$ respectively.}
\end{sidewaysfigure}

The relaxation of the $f_{0}$ distribution function component are
compared for the gas and liquid phases at three different times in
Figure \ref{fig:ContourArg}. At $t^{*}=1$, there are only small
differences between the contours reflecting similar energy relaxation
rates between the two phases initially. At $t^{*}=10$, a bulge is
beginning to develop in the gas-phase contour in the energy region
between $0.1-0.5$ eV, which corresponds to the Ramsauer minimum in
the gas-phase momentum transfer cross-section. In this region, the
gas-phase momentum transfer cross-section dips below the liquid cross-section,
which has resulted in this enhancement of the diffusive flux in this
range. At higher energies the liquid cross-section is less than the
gas-phase cross-section, which has resulted in enhanced diffusive
flux. At $t^{*}=100$ these effects are even more pronounced. 

In Figure \ref{fig:ContourArg} the $f_{1}$ component contours for
the gas and liquid phases of argon are compared for the same three
times. At the first time, $t^{*}=1$, there is already significant
differences in the $f_{1}$ contours, with the largest change occurring
in the Ramsauer minimum range in the gas-phase case. This highlights
again the difference in the timescales of the energy and momentum
relaxation between the two phases. As time increases, greater differences
develop between the $f_{1}$ contours particularly around the Ramsauer
minimum and at the high energy range for the reasons previously discussed.

The number density as a function of time is shown in Figure \ref{fig:ContourArg}.
The behavior of the number density profiles is consistent with the
behaviour of the $f_{0}$ and $f_{1}$ profiles. At $t^{*}=1$ there
is no noticeable difference in the two number density profiles. At
later times it is clear that, despite the Ramsauer minimum in the
gas-phase, the liquid-phase experiences the greater diffusion rate
overall. For the electric field considered and initial source distribution,
the average drift velocity for both the gas and liquid phases is small
compared to the diffusion rates. 

\begin{figure}[H]
\begin{minipage}[t]{1\columnwidth}%
\subfloat{\includegraphics[width=0.33\textwidth]{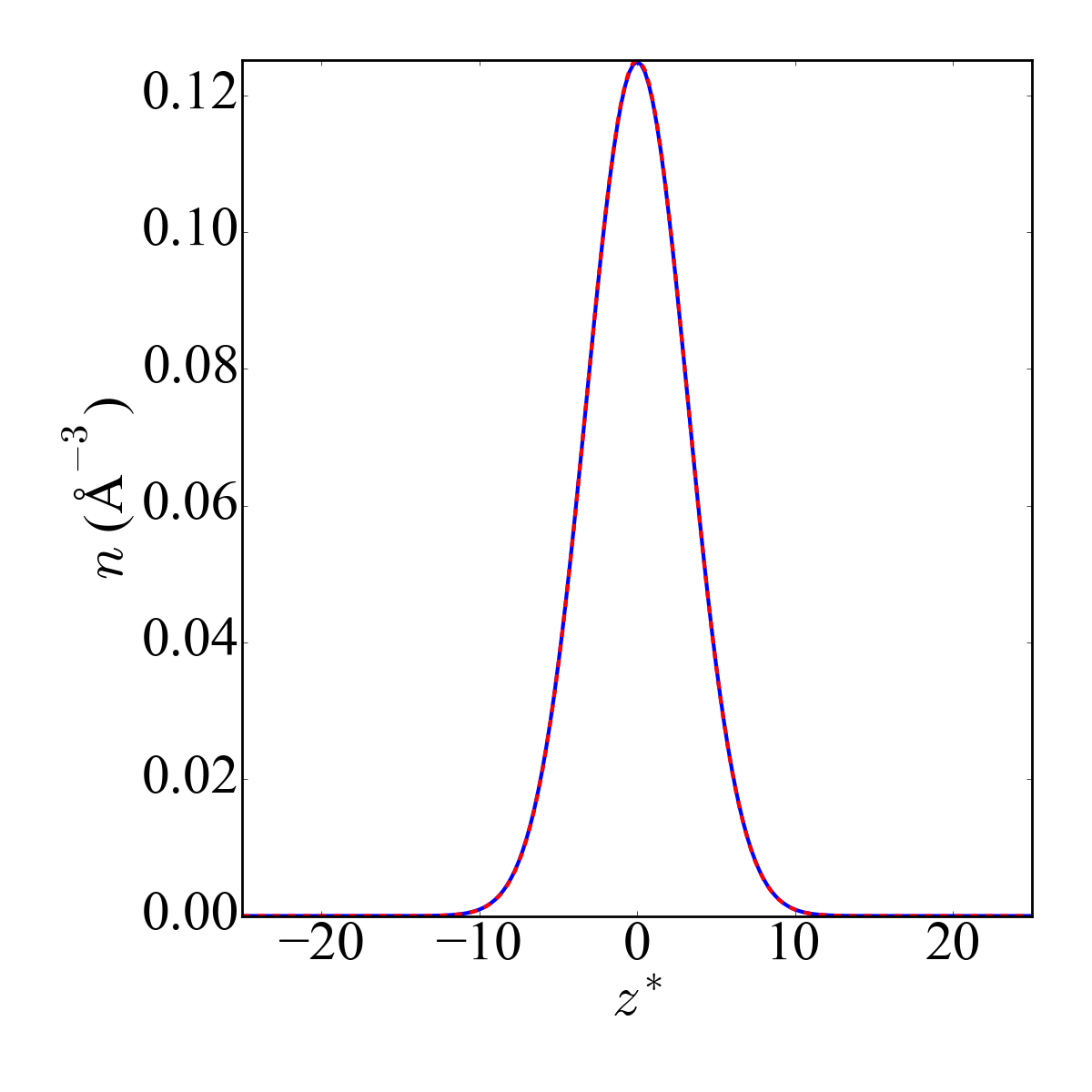}}\subfloat{\includegraphics[width=0.33\textwidth]{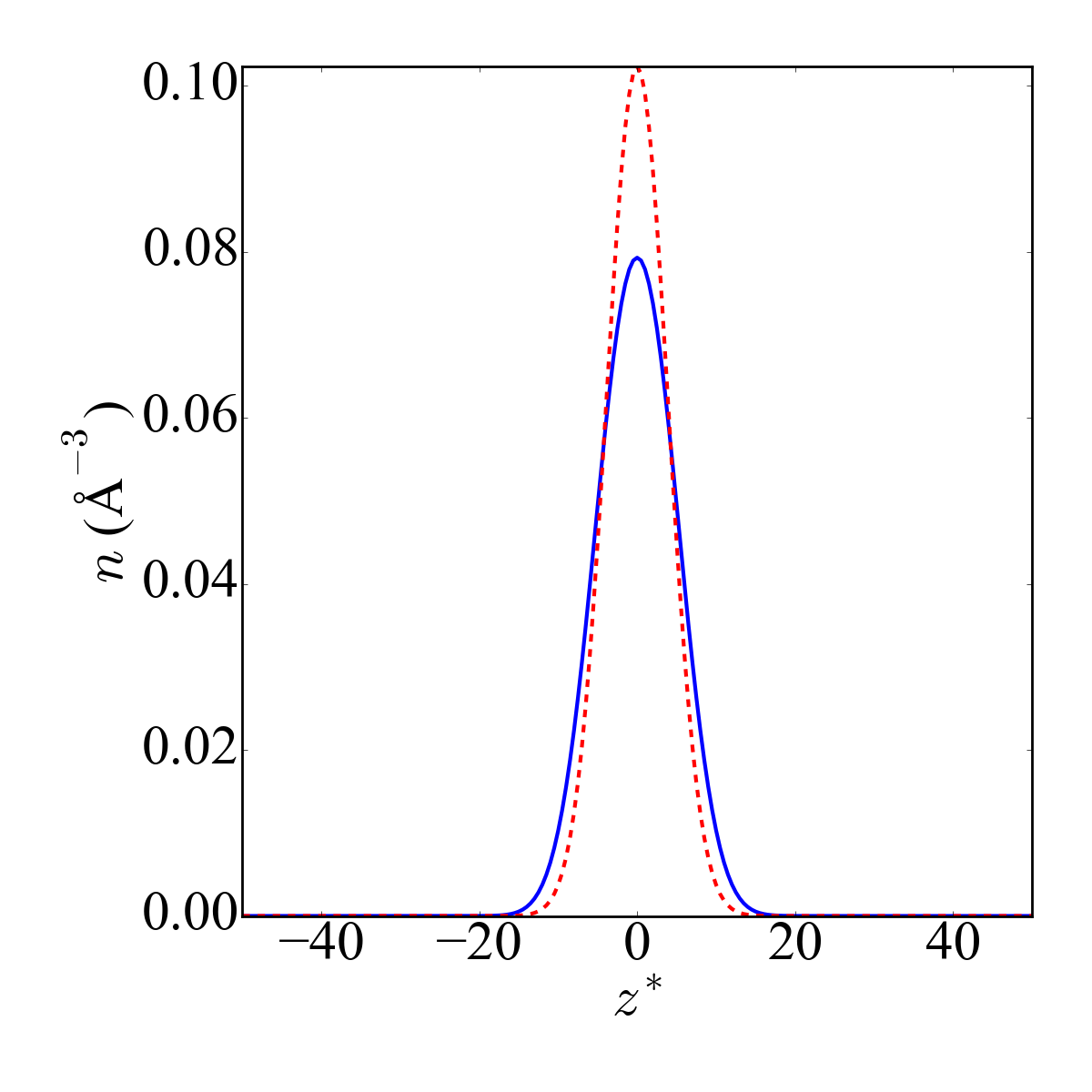}}\subfloat{\includegraphics[width=0.33\textwidth]{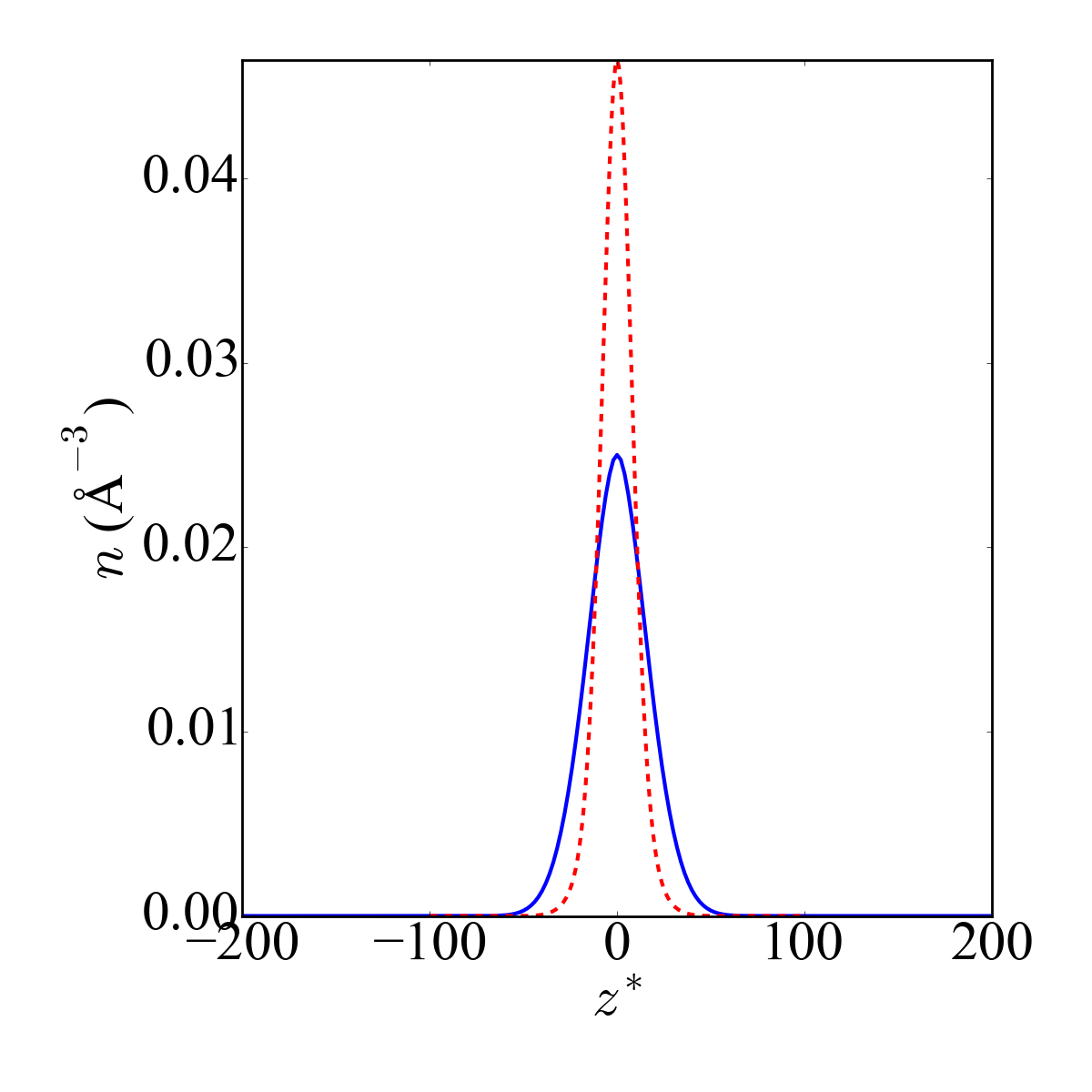}}%
\end{minipage}

\protect\caption{\label{fig:nDArg}Temporal evolution of the number density profiles
for gas-phase (dashed lines) and liquid-phase (solid lines) argon.
The three columns represent the times, $t^{*}=1,\ 10,\mbox{ and }100$
respectively.}
\end{figure}

\section{Conclusion}

We have developed a full multi-term, space-time dependent solution
of the electron Boltzmann equation in gases and liquids capable of
modeling non-hydrodynamic conditions. The flexibility of the algorithm
lies in solving the Boltzmann equation's Green's function, knowledge
of which allows one to construct the solution for other experimental
configurations e.g. the SST experiment and similar applications. Operator
splitting has been employed to efficiently evolve the energy-space
and configuration-space components individually with tailored numerical
schemes.

The theory and associated code was first applied to a simple hard-sphere
benchmark model liquid, where structure effects were simulated by
the Percus-Yevick structure factor as a function of the volume fraction,
$\Phi$. The inclusion of an inelastic channel was a key test of the
algorithm's ability to reproduce non-hydrodynamic phenomena. Periodic
spatial structures developed in the space-time and steady-state profiles
for the distribution function components and associated transport
properties, the periodicity of which is directly related to the threshold
energy of the inelastic process. We observed that these periodic structures
arose on shorter times scales when coherent scattering effects became
important. The steady-state profiles constructed for various volume
fractions also reproduced the non-hydrodynamic oscillatory structures
expected. The asymptotic transport coefficients calculated from the
non-hydrodynamic solution of Boltzmann's equation were also shown
to be consistent with the values calculated from a hydrodynamic solution
of Boltzmann's equation.

Finally, the cross-sections calculated in \cite{Boyletal15} were
used to investigate the spatio-temporal evolution of electrons in
gas-phase and liquid-phase argon. The two momentum-transfer cross-sections
feature different qualitative and quantitative behaviours. Striking
differences in the evolution of the components of the phase-space
distribution were apparent, reflecting the differences in the gas-phase
and liquid-phase cross-sections, particularly the absence of a Ramsauer
minimum in the liquid-phase. This highlights the problems associated
with treating liquid systems as gaseous systems with increased density,
which has implications for various applications including liquid argon
time projection chambers.
\begin{acknowledgments}
This work was supported under the Australian Research Council's (ARC)
Centre of Excellence and Discovery programs. 
\end{acknowledgments}
\bibliographystyle{apsrev}
\bibliography{SpaceGasLiq,library}

\end{document}